\documentclass[journal]{IEEEtran}
\usepackage[caption=false]{subfig}
\usepackage{url}
\usepackage{graphicx}
\usepackage{amsfonts}
\usepackage{hyperref}
\usepackage{balance}
\usepackage{makecell}
\usepackage{stfloats}
\usepackage{color,soul}
\usepackage{amsmath}
\usepackage{multirow}
\usepackage{hhline}
\usepackage{cite}
\usepackage{amssymb}
\usepackage{amsthm}
\usepackage{mathtools}
\usepackage{multicol}
\usepackage{comment}
\usepackage{adjustbox}
\usepackage{xcolor,colortbl}
\usepackage{enumitem}
\usepackage{blindtext}
\definecolor{Gray}{gray}{0.9}
\usepackage{tikz}
\newcolumntype{M}[1]{>{\centering\arraybackslash}m{#1}}

\begin{document}

\title{Aerial Computing: A New Computing Paradigm, Applications, and Challenges}

\author{Quoc-Viet~Pham,~\IEEEmembership{Member,~IEEE},
Rukhsana~Ruby,~\IEEEmembership{Member,~IEEE},
Fang~Fang,~\IEEEmembership{Member,~IEEE},\\
Dinh~C.~Nguyen,~\IEEEmembership{Member,~IEEE},
Zhaohui~Yang,~\IEEEmembership{Member,~IEEE},
Mai~Le,\\
Zhiguo~Ding,~\IEEEmembership{Fellow,~IEEE}
and Won-Joo~Hwang,~\IEEEmembership{Senior Member,~IEEE}

\thanks{Quoc-Viet~Pham is with the Korean Southeast Center for the 4th Industrial Revolution Leader Education, Pusan National University, Busan 46241, Korea.}
\thanks{Rukhsana~Ruby is with College of Computer Science and Software Engineering, Shenzhen University, China.}
\thanks{Fang~Fang is with the Department of Electrical and Computer Engineering and the Department of Computer Science, Western University, London, ON N6A 3K7, Canada.}
\thanks{Dinh~C.~Nguyen is with the School of Electrical and Computer Engineering, Purdue University, USA.}
\thanks{Zhaohui~Yang is with the Centre for Telecommunications Research, Department of Engineering, King’s College London, WC2R 2LS, UK.}
\thanks{Mai~Le is with the Department of Information Convergence Engineering, Pusan National University, Busan 46241, Korea.}
\thanks{Zhiguo~Ding is with the School of Electrical and Electronics Engineering, The University of Manchester, Manchester M13 9PL, UK.}
\thanks{Won-Joo~Hwang is with the Department of Biomedical Convergence Engineering, Pusan National University, Yangsan 50612, Korea.}
\thanks{E-mails: vietpq@pusan.ac.kr, ruby@szu.edu.cn, fang.fang@uwo.ca, nguye772@purdue.edu, yang.zhaohui@kcl.ac.uk, maile2108@gmail.com, zhiguo.ding@manchester.ac.uk, wjhwang@pusan.ac.kr.}
}

\maketitle

\begin{abstract}
In existing computing systems, such as edge computing and cloud computing, several emerging applications and practical scenarios are mostly unavailable or only partially implemented. To overcome the limitations that restrict such applications, the development of a comprehensive computing paradigm has garnered attention in both academia and industry. However, a gap exists in the literature owing to the scarce research, and a comprehensive computing paradigm is yet to be systematically designed and reviewed. This study introduces a novel concept, called \textit{aerial computing}, via the amalgamation of aerial radio access networks and edge computing, which attempts to bridge the gap. Specifically, first, we propose a novel comprehensive computing architecture that is composed of low-altitude computing, high-altitude computing, and satellite computing platforms, along with conventional computing systems. We determine that aerial computing offers several desirable attributes: global computing service, better mobility, higher scalability and availability, and simultaneity. Second, we comprehensively discuss key technologies that facilitate aerial computing, including energy refilling, edge computing, network softwarization, frequency spectrum, multi-access techniques, artificial intelligence, and big data. In addition, we discuss vertical domain applications (e.g., smart cities, smart vehicles, smart factories, and smart grids) supported by aerial computing. Finally, we highlight several challenges that need to be addressed and their possible solutions.
\end{abstract}


\IEEEpeerreviewmaketitle
\section{Introduction}
\label{sec:Introduction}
Research communities have started to define roadmaps for sixth-generation (6G) networks \cite{tataria20216g}. 
As various Internet-of-Things (IoT) applications are set to emerge in 6G, additional data need to be collected and transmitted. However, IoT devices are constrained by factors, such as batteries, transmission power, and processing capacity. To facilitate data transmission from IoT devices worldwide, several studies have investigated aerial access networks (AANs), which can offer line-of-sight (LoS) communication links, favorable channels, and improved coverage. 
An AAN generally consists of three primary components: low-altitude platforms (LAPs), high-altitude platforms (HAPs), and low-Earth-orbit (LEO) satellites \cite{pham2022aerial}. These platforms fully complement the conventional terrestrial access network to create a future access network in 6G that can provide wireless services with global coverage and diverse quality of service (QoS). Moreover, to better support emerging compute-intensive applications, such as fully autonomous vehicles, flying taxis, holographic communications, and virtual reality (VR), edge computing (e.g., fog computing and multi-access edge computing (MEC)) is a promising concept owing to the availability of powerful computing and storage resources at the edge of the network \cite{jiang2021ai}. 
The amalgamation of AANs and edge computing introduces a novel concept referred to as \textit{aerial computing}. Aerial computing is expected to provide advanced services, such as communication, computing, caching, sensing, navigation, and control, at a global scale.

Aerial computing is considered a promising paradigm that enables local data analysis and real-time service provisioning in the air. Facilitated by advantages, such as high mobility, fast deployment, global availability, scalability, and flexibility, aerial computing complements conventional computing paradigms (e.g., cloud computing, fog computing, and MEC) and is thus considered a pillar of the comprehensive computing infrastructure in future 6G networks \cite{ke2021edge}. 
For example, in \cite{yu2020joint}, IoT data over small areas could be collected and processed cooperatively by a set of unmanned aerial vehicles (UAVs) at low altitudes, whereas IoT data over large-scale areas could be executed by HAPs at high altitudes. 
In \cite{garg2018uav}, UAVs were leveraged as intermediate points between vehicles and the network edge to enable data collection and task preprocessing and thus facilitate data transfer in smart vertical domain environments. 
Meanwhile, in the race toward global Internet connectivity among tech giants, such as SpaceX, Google, Amazon, and Facebook, several LEO satellite systems have been deployed to provide Internet services to users, particularly in rural and hard-to-reach areas, with a performance that is comparable to that of the current terrestrial mobile network \cite{satellite_race}. However, these systems have not yet been integrated into current computing systems as key components. 

The above observations emphasize the importance of a comprehensive computing architecture for future 6G systems. As a native constituent, aerial computing is composed of low-altitude and high-altitude computing platforms as well as satellite computing platforms. It can carry its full complement of conventional terrestrial computing paradigms. Conceptually, aerial computing can facilitate various smart industrial applications, such as smart cities, smart vehicles, smart factories, and smart grids. Moreover, it is expected to provide a reference 6G computing architecture for future studies. 

\subsection{Visions Toward a Comprehensive Computing Infrastructure}
\label{sec:Introduction_Vision}
In recent years, several surveys on edge computing and mobile networks have been conducted. For example, enabling MEC technologies, including virtual machines (VMs), network function virtualization (NFV), software-defined networking (SDN), and network slicing were reviewed in \cite{taleb2017multi}. The use of MEC for IoT applications, such as critical and massive IoT, wearable IoT, smart energy, and IoT automotive, was discussed in \cite{porambage2018survey}. 
Computation offloading, resource management, and optimal placement problems in edge computing systems were discussed in \cite{mach2017mobile, mao2017survey, sonkoly2021survey}. The amalgamation of MEC and AI, referred to as edge intelligence and intelligent edge, has been reviewed in several studies \cite{wang2020convergence, xu2021edge}. Further, certain reviews focused on the integration of edge computing with emerging technologies, such as blockchain \cite{yang2019integrated} and fifth-generation (5G) and beyond networks \cite{pham2020survey}. More recently, privacy/security issues and solutions of MEC services were reviewed \cite{danaweera2021Survey}. 

MEC is a key enabler of many wireless services offering better QoS. Moreover, MEC has the potential to be closely integrated with emerging technologies and systems \cite{yang2019integrated, pham2020survey}. In addition, these studies elaborate on the various challenges of MEC that need to be examined in beyond-5G networks, such as resource optimization, security and privacy, and real-time implementation. With the expected extraordinary popularization of the next-generation Internet and the emergence of killer applications, such as VR, space tourism, fully autonomous vehicles, holographic communications, and deep-sea sight, the current MEC infrastructure needs to be further expanded both vertically and horizontally to complement existing edge computing systems. However, the aforementioned reviews 
\cite{taleb2017multi, porambage2018survey, mao2017survey, mach2017mobile, sonkoly2021survey, wang2020convergence, xu2021edge, yang2019integrated, pham2020survey, danaweera2021Survey} 
primarily focus on 5G network scenarios, technologies, and applications, but the new requirements for a comprehensive 6G computing infrastructure have not been reviewed yet. 

Future 6G systems have been envisioned in recent studies. For example, the authors of \cite{tariq2020speculative} discussed their vision for use cases, enabling technologies, and overcoming the challenges of 6G networks. In \cite{chen2020vision}, a set of key performance indicators (KPIs) were defined for 6G, such as peak data rate $>$100 Gbps, traffic density $>$100 Tb/s/km$^2$, network density $>$10 million connections per km$^2$, mobility $>$1000 km/h, centimeter-level positioning accuracy, reliability $>$99.999\%, receiver sensitivity $<$-130 dBm, $3$-times spectral efficiency, and $10$-times energy efficiency. Six technical trends were discussed in \cite{chen2020vision}, namely, coverage expansion, new resources and bandwidth spectrum, new modulation techniques, enhancing the system capacity, adding computing and intelligence capabilities, and three-dimensional (3D) network architecture. Regarding 3D networks, several studies have reviewed UAV communications \cite{fotouhi2019survey}, HAP networks \cite{kurt2021vision}, airborne communications \cite{cao2018airborne}, satellite communications \cite{kodheli2020satellite}, terrestrial-satellite integrated networks \cite{xie2020satellite}, and future aerial networks \cite{baltaci2021survey}. Recently, the primary aspects of aerial radio access networks, such as architectures, network design, enabling technologies, and applications, were reviewed in \cite{dao2021aran_Survey}. Owing to excellent features and numerous applications, many researchers believe that AI techniques are a key component of many 6G systems. For example, machine learning (ML) is considered a double-edged sword for privacy in 6G \cite{sun2020machine}.
Moreover, AI is an integral part of future networks, such as sensing AI at receivers, on-device AI, access AI (i.e., transmission at PHY, MAC, and network layers), and data-provenance AI \cite{nguyen2020enabling}.

The above surveys mainly concentrate on the KPIs, enabling technologies, potential applications, and major challenges of 6G. The existing reviews emphasize the importance of a comprehensive computing infrastructure for the foundation of 6G in the next 10 years. However, despite promising studies and an urgent need for a comprehensive computing infrastructure, a large gap exists in the existing studies because aerial computing has not been systematically reviewed. This paper attempts to bridge this gap by introducing a new computing architecture that can be employed in future network systems. This paper also presents a comprehensive review of aerial computing. Table~\ref{Table:Summary_ExistingSurveys} summarizes the existing surveys on edge computing and aerial communications. It also highlights the contributions of the present study. 

\begin{table*}[ht!]
    \renewcommand{\arraystretch}{1.15}
	\caption{Summary of existing surveys on edge computing and future wireless networks. 
	}
	\label{Table:Summary_ExistingSurveys}
	\centering
	\begin{tabular}{|c|c|c|c|p{7.20cm}|p{5.0cm}|}
		\hline 
		\multirow{2}{*}{\textbf{Reference}}  & \multicolumn{3}{c|}{\textbf{Main theme}} & \multirow{2}{*}{\textbf{Key Contributions}} & \multirow{2}{*}{\textbf{Limitations}}  \\ 
		
		\cline{2-4}
		{} & \textbf{6G/AI} & \textbf{MEC} & \textbf{AAN} & {} & \\
		\hline
		\hline
		
		\multirow{1}{*}{\cite{taleb2017multi}} 
		&  
		& \multirow{1}{*}{\checkmark} &  &  Enabling technologies of MEC systems, such as VM, containers, NFV, SDN, and network slicing, are reviewed. 
		& This paper does not focus on aerial computing and its role.
		\\ \hline
		
		\multirow{1}{*}{\cite{porambage2018survey}} 
		&  
		& \multirow{1}{*}{\checkmark} &  &  MEC applications for IoT service realization in 5G networks.
		& The role and applications of aerial computing have not been presented.
		\\ \hline

		\multirow{1}{*}{\cite{mao2017survey, mach2017mobile, sonkoly2021survey}} 
		&  
		& \multirow{1}{*}{\checkmark} &  &  Reviews of seminal MEC architectures, computation offloading issues, resource management, and optimal placement problems. 
		& The role and applications of aerial computing are not discussed.
		\\ \hline
		
		\multirow{1}{*}{\cite{wang2020convergence, xu2021edge}} 
		&  
		& \multirow{1}{*}{\checkmark} &  &  Interactions between AI techniques and MEC systems; that is, intelligent edge and edge intelligence. 
		& The role of aerial computing in enabling a comprehensive computing infrastructure in 6G is not considered.
		\\ \hline
		
		\multirow{1}{*}{\cite{yang2019integrated}} 
		&  
		& \multirow{1}{*}{\checkmark} &  &  Integration of blockchain and MEC systems. 
		& Only the integration of blockchain into edge computing systems is presented.
		\\ \hline
		
		\multirow{2}{*}{\cite{pham2020survey}} 
		&  
		& \multirow{2}{*}{\checkmark} &  &  Integration of MEC with 5G technologies, such as cloud radio access networks, massive multi-input multi-output (MIMO), unmanned aerial vehicle (UAV) communications, and AI. 
		& Only the integration of MEC with 5G technologies is reviewed, while the role and applications of aerial computing are ignored.
		\\ \hline
		
		\multirow{1}{*}{\cite{danaweera2021Survey}} 
		&  
		& \multirow{1}{*}{\checkmark} &  &  Comprehensive survey on privacy/security issues and solutions in MEC systems. 
		& The paper reviews only the security and privacy aspects of MEC systems.
		\\ \hline
		
		\multirow{1}{*}{\cite{tariq2020speculative}} 
		& \multirow{1}{*}{\checkmark} 
		&  &  &  A speculative study on promising applications and major technologies in future 6G networks. 
		& Only the vision of future 6G networks is discussed.
		\\ \hline
	
	    \multirow{1}{*}{\cite{chen2020vision}} 
	    & \multirow{1}{*}{\checkmark} 
	    &  &  &  A vision of KPIs, technologies, technical trends, and challenges of 6G networks. 
	    & Only a brief discussion of future 6G networks is presented.
	    \\ \hline
	    
	    \multirow{2}{*}{\cite{kodheli2020satellite, xie2020satellite, fotouhi2019survey, kurt2021vision, cao2018airborne, baltaci2021survey}} 
	    &  
	    &  & \multirow{2}{*}{\checkmark} & 3D networking, such as UAV communications~\cite{fotouhi2019survey}, HAP systems \cite{kurt2021vision}, airborne communications \cite{cao2018airborne}, satellite communications \cite{kodheli2020satellite}, terrestrial-satellite integrated networks \cite{xie2020satellite}, and wireless communications for future AANs \cite{baltaci2021survey}.
	    & These papers primarily focus on the communication aspects of AAN, while the computing aspects and the role of aerial computing are ignored.  
	    \\ \hline
		
		\multirow{2}{*}{\cite{dao2021aran_Survey}} 
		& \multirow{2}{*}{\checkmark} 
		&  
		& \multirow{2}{*}{\checkmark} & Technical aspects of AANs in the 6G context and key enablers.
		& The computing aspects and applications of aerial computing are not presented. 
		\\ \hline
		
        \multirow{1}{*}{\cite{sun2020machine}} 
		& \multirow{1}{*}{\checkmark} 
		&  
		&  & A comprehensive survey of the alliance between ML and 6G privacy.
		& The paper focuses only on privacy aspects and its alliance with ML.
		\\ \hline
		 
		\multirow{2}{*}{\cite{nguyen2020enabling}} 
		& \multirow{2}{*}{\checkmark} 
		&  
		&  &  Four important domains of AI in 6G are reviewed, namely, sensing AI, on-device AI, access AI, and data-provenance AI. 
		& The paper focuses on the role and applications of AI in future networks.
		\\ \hline
		 
		\multirow{2}{*}{This paper} 
		& \multirow{2}{*}{\checkmark} 
		& \multirow{2}{*}{\checkmark} & \multirow{2}{*}{\checkmark} &  A comprehensive survey of aerial computing. In particular,
		\begin{itemize}
		    \item We introduce a novel architecture of aerial computing, which complements the existing computing infrastructures. Important features of aerial computing, such as mobility, availability, scalability, flexibility, and simultaneity, have also been analyzed in detail.
		    
		    \item We provide in-depth discussions of key enablers of aerial computing and the use of aerial computing in important vertical applications, including smart cities, smart vehicles, smart factories, and smart grids. 
		    
		    \item Key challenges of aerial computing are presented along with potential solutions and future research directives. 
		\end{itemize}
		 & 
		\\ \hline
	\end{tabular}
\end{table*}

\subsection{Contributions and Research Methodologies}
\label{sec:Introduction_Contributions}
As discussed above, aerial computing is a promising concept for providing computing resources and wireless services on a global scale. This paper attempts to bridge the gap in the existing literature by presenting an in-depth review of aerial computing with a vision toward a comprehensive computing infrastructure in future 6G networks. 
The review covers the following aspects.

\begin{figure*}[t]
\centering
\includegraphics[width=0.985\linewidth]{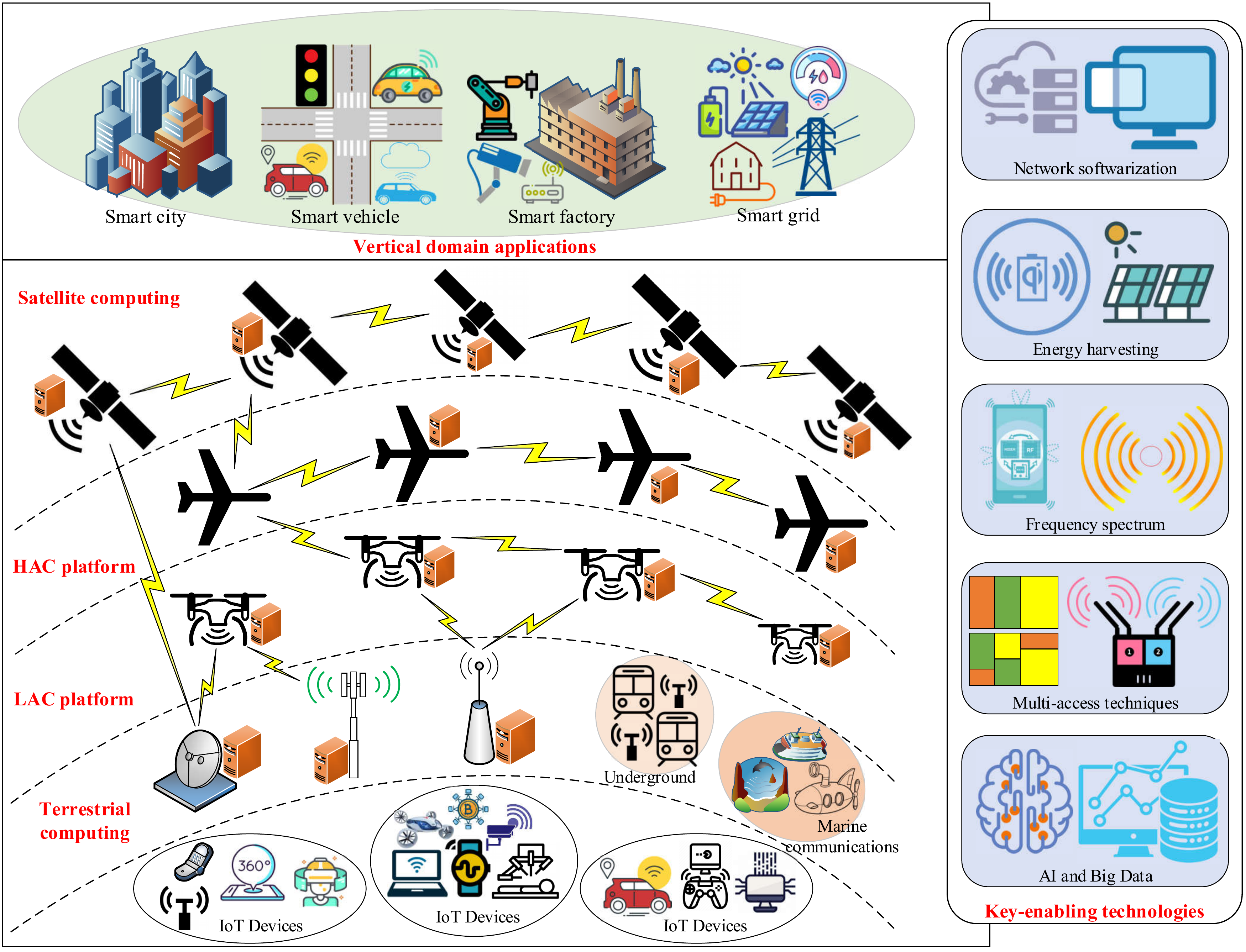}
\caption{A novel comprehensive 6G computing architecture.} 
\label{Fig:AC_Architecture}
\end{figure*}

\textit{\textbf{Fundamentals and Designs}} (Section~\ref{sec:Design}): First, to enrich the full understanding of the concept of aerial computing, we present the system architecture, which is illustrated in Fig.~\ref{Fig:AC_Architecture}. This comprehensive computing architecture is expected to be the key enabler of many computationally intensive applications that will be available in 6G systems. Following the analysis of the proposed architecture, we elaborate upon the important features of aerial computing, including ubiquity, mobility, availability, simultaneity, and scalability, and further differentiate between aerial computing from other computing paradigms. Further, we present important design problems for an aerial computing system, including computation, communication, energy consumption, and latency models. 
    
\textit{\textbf{Enabling Technologies}} (Section~\ref{sec:Technologies}): To realize aerial computing in 6G systems, we discuss a set of key enabling technologies, which include {network softwarization, energy refilling, frequency spectrum, multi-access techniques, and AI and big data}. In particular, energy refilling is crucial for maintaining the sustainable operations of aerial components. With regard to the operational management of aerial computing systems, we review state-of-the-art studies on softwarization techniques, including VM, network slicing, NFV, and SDN. These softwarization techniques are important for facilitating flexible and programmable operations and the deployment acceleration of aerial computing systems. In addition, we discuss frequency spectrum and multi-access techniques (e.g., massive MIMO, intelligent reflecting surface (IRS), and non-orthogonal multiple access (NOMA)), used in aerial computing systems. Finally, AI and big data analytics are considered important technologies for exploiting massive data generated from aerial computing, thus improving system performance.
    
\textit{\textbf{Vertical Domain Applications}} (Section~\ref{sec:Applications}): The integration of wireless technologies into industrial automation systems has shown great potential in creating more efficient and productive vertical domain applications (i.e., smart cities, smart vehicles, smart factories, and smart grids). Further, we present an extensive review of state-of-the-art studies on these vertical domain applications in the context of 6G aerial computing. In this regard, we discuss the manner in which aerial computing can support vertical domain applications. 
    
\textit{\textbf{Research Challenges and Potential Directions}} (Section~\ref{sec:Challenges}): To further drive research into aerial computing in the future, we discuss the major challenges, possible solutions, and potential directives. Aerial computing systems face critical challenges, such as energy efficiency, efficient resource management, and network stability. Further, large-scale network optimization is another critical issue for embedding distributed and learning approaches such that aerial computing systems can efficiently operate on a large scale. Finally, security, privacy, and trust are important aspects of aerial computing but have not been adequately addressed by existing studies. 

\section{Fundamentals and Design}
\label{sec:Design}
In this section, we present the system architecture and fundamental features of aerial computing.

\subsection{System Architecture}
\label{sec:Design_Architecture}
As shown in Fig.~\ref{Fig:AC_Architecture}, we propose a novel aerial computing-based 6G architecture wherein the aerial components are positioned in a hierarchical manner. In addition to conventional computing infrastructure, the proposed aerial computing architecture consists of four main entities: IoT devices, low-altitude computing (LAC) platforms, high-altitude computing (HAC) platforms, and satellite computing platforms, along with vertical domain applications and key enabling technologies.

    \textit{\textbf{IoT}}: This layer refers to any IoT device that generates data and comprises compute-intensive applications that need to be executed by aerial computing systems. For example, wearable devices, such as smart watches, eyeglasses, fitness trackers, and body-mounted sensors are responsible for monitoring and collecting the health information (e.g., heart rate, physical activity, and blood pressure). These devices may offload collected data to an edge node for further processing because they are typically resource-constrained and have a relatively small size. Aerial components (e.g., UAVs and drones) can be a part of this IoT layer when they are dispatched to collect data from ground sensors. To maximize the flight time and data collection efficiency, the UAV offloads the collected data to a terrestrial MEC server and/or other more powerful UAVs. 
    
    \textit{\textbf{Terrestrial Computing}}: This layer comprises conventional terrestrial computing paradigms, such as fog computing, MEC, and cloudlet. Typically, computing nodes are deployed at preset locations. According to \cite{pham2020survey}, the potential deployment locations of an MEC server include routers, IoT gateways, macro base stations (BSs), small cell BSs, optical network units, radio network controller sites, and wireless fidelity (WiFi) access points, whereas the common locations of a fog node are gateways, intermediate nodes (between the cloud and end devices), and network elements (e.g., routers and switches) \cite{mukherjee2018survey}. The collaborative and hierarchical MEC framework utilizes the advantage of many applications and services, as edge devices with different capabilities can collaboratively perform data processing and task execution \cite{wu2020toward}.
    
    \textit{\textbf{Low-Altitude Computing}}: The LAC platform is positioned in the lowest tier of aerial computing systems with an altitude of 0--10 km above the Earth's surface. This platform offers several advantages, such as cost-effectiveness, rapid deployment, and LoS communication links. Hence, LAC systems are highly effective for emergency scenarios, military surveillance, temporary events, and IoT data collection and processing \cite{tran2022uav}. 
    The authors of \cite{hu2020wireless} demonstrated that the UAV can be leveraged for data collection, aerial relaying, and aerial computing of the tasks offloaded from ground IoT devices. The primary components of the LAC platform are drones and UAVs equipped with computing capabilities, which typically have limited battery capacities, low computing resources, and low endurance. Consequently, several solutions have been proposed to address these challenges, such as wireless- and solar-powered UAV-MEC and task offloading to more powerful terrestrial and HAC-MEC servers \cite{liu2019minimization}. There exist two types of LAC servers: fixed-wing and rotary-wing. The former does not hold up in the air and needs to move forward continuously, whereas the latter can take off and land vertically.
    
    \textit{\textbf{High-Altitude Computing}}: The HAC platform comprises both manned and unmanned aerial components (e.g., airplanes, airships, and balloons) that operate at an altitude ranging from 17 to 50 km. The advantages of HAC systems over other platforms include large-area coverage with the cell size of up to 10 km, adaptability to traffic demands, rapid deployment compared to terrestrial computing and satellite computing, high endurance compared to LAC platforms, and green operation capability with solar power \cite{tozer2001high}. At the World Radio Conference (WRC)-19, several frequency bands were allocated to HAC systems by the ITU Radio communication Sector, including the 31--31.3 and 38--39.5 GHz bands for worldwide use, 47.2--47.5 and 47.9--48.2 GHz for administration use, and 21.4--22 and 24.25--27.5 GHz for fixed services in ITU Regions 2 \cite{HAPS_fbands}. In recent years, several industrial projects have been observed in LAC platforms for wireless and computing services, such as flying cells on wings \cite{COWs}, Google Skybender \cite{Skybender}, and ApusDuo \cite{ApusDuo}. Owing to the quasi-stationary feature, HAC platforms can form a collaborative HAC-MEC server with more powerful capabilities and higher time endurance, aiding in the execution of the computation tasks that are offloaded from LAC and/or satellite systems. 
    
    \textit{\textbf{Satellite Computing}}: This layer is composed of LEO satellites with computing capabilities. LEO satellites typically operate at an altitude of approximately 80--2000 km above the Earth's surface. For example, SpaceX launched 60 LEO satellites at altitudes not greater than 580 km in May 2019. These satellites can provide Internet services globally with a data rate ranging from 50 to 150 Mbps and a latency ranging from 20 to 40 ms (\url{www.starlink.com/}). 
    Amazon also initiated a satellite project called Kuiper, which plans to launch a constellation of 3200 satellites at an altitude of approximately 610 km \cite{AmazonKuiper}. These LEO satellites are expected to provide various wireless and computing services, particularly when other computing platforms are temporarily unavailable and/or satellite computing is the only option. Unlike in LAC and HAC platforms, users are not usually connected with satellite computing systems because of the excessive hardware cost incurred when MEC servers are embedded in LEO satellites. Instead, LEO satellites can act as relays to receive and transmit computation tasks/results between users and MEC servers. However, computing directly at the satellite edge server is beneficial for applications with sparse users; that is, users are located at different geographical locations \cite{zhang2019satellite, xie2020satellite}.
    
    \textit{\textbf{Industrial Applications}}: Aerial computing supports and enables (new) industrial applications owing to its distinctive features, such as high mobility, always-available computing, and scalability (discussed in Section~\ref{sec:Design_Features}). For example, aerial computing systems can support large-scale vehicle systems and/or advanced traffic signal infrastructures to improve the experience of in-car infotainment applications and reduce traffic congestion by deploying massive MEC on roadside units, power systems, buildings, and highway infrastructure. In Section~\ref{sec:Applications}, we comprehensively discuss the use of aerial computing for four primary industrial applications: smart cities, smart vehicles, smart factories, and smart grids. 
    
    \textit{\textbf{Enabling Technologies}}: The realization of aerial computing is facilitated by several key enabling technologies, including {network softwarization, energy refilling, frequency spectrum, multi-access techniques, and AI and big data}. A thorough discussion of these technologies in the context of aerial computing is presented in Section~\ref{sec:Technologies}.


\subsection{Desirable Features}
\label{sec:Design_Features}
Aerial computing retains the base features of the concept of edge computing, including on-premises, proximity, lower latency, location awareness, and network contextual information. In addition, it has certain fundamental features that differentiate it from other computing paradigms (e.g., fog computing and cloudlet), which are discussed as follows. 
\begin{itemize}
    \item \textit{Mobility}: Aerial platforms can be deployed quickly to support on-demand computing applications. While LAC platforms can enable the rapid roll-out of computing services, the deployment of a terrestrial computing system is time-consuming owing to planning procedures and civil works involved. From the perspective of the user, the overlaying architecture of LAC, HAC, and satellite computing helps aerial computing systems flexibly satisfy different computation tasks from end IoT devices. 
    
    \item \textit{Availability}: Aerial computing systems are usually available as they are not much affected by natural disasters, unlike terrestrial computing, and can be easily and quickly deployed for particular computation tasks. The availability of aerial computing also relies on its hierarchical architecture with different computing platforms at different altitudes and coverage distances.

    \item \textit{Scalability}: Aerial computing systems can provide on-demand computing services as well as applications with geographically distributed users. As the aerial components can collaboratively form computing clusters, massive IoT devices can be served, thereby improving user satisfaction and guaranteeing service continuity. 
    
    \item \textit{Flexibility}: Unlike terrestrial computing, wherein MEC and fog nodes are typically deployed at pre-specified locations, aerial computing entities can be changed easily to flexibly suit different situations. Further, enhanced computing services may be provided via additional aerial computing platforms, such as LAC and HAC.

    \item \textit{Simultaneity}: Different aerial computing platforms can provide computing services to geographical users simultaneously. Thus, users in different countries can request computing services to LEO satellite computing services concurrently. Therefore, aerial computing is regarded as a full complement of terrestrial computing infrastructure, which is usually used for localized computing services. 
\end{itemize}

\begin{table*}[t]
    \renewcommand{\arraystretch}{1.205}
    \caption{Comparison of aerial computing with other edge computing paradigms.}
    \centering
    \begin{tabular}{|p{1.5cm}|p{3.56cm}|p{3.56cm}|p{3.56cm}|p{3.56cm}|}
        \cline{2-5}
        \multicolumn{1}{c|}{} & \textbf{Cloudlet} & \textbf{Fog Computing} & \textbf{MEC} & \textbf{Aerial Computing} \textbf{} \\
        \hline
        {Introduced by} & {Satyanarayanan, 2009 \cite{satyanarayanan2009case}} & {Cisco, 2012} & {ETSI, 2014} & {This work, 2021} 
        \\ \hline
        
        Purpose & \multicolumn{3}{c|}{Moving cloud computing capabilities to the network edge} & Making computing capabilities available at all times, at both the network edge and air
        \\ \hline
        
        {Deployment} & {Business premises (e.g., shopping malls, companies, and personal computers)} & {Strategic locations both indoors and outdoors (e.g., IoT gateways, routers, and switches)} & {Various locations within radio access networks (RANs) (e.g., radio towers and gateways)} & {Locations in the air and within RANs (e.g., radio towers, UAVs, and LEO satellites)} 
        \\ \hline
        
        {Application \newline Examples} & {Locally resource-intensive and interactive applications} & {Smart vertical domain applications (e.g., smart cities and smart health) and video surveillance} & {Content caching, autonomous vehicle, augmented reality (AR), data analytics} & {On-demand applications and applications with sparse user distribution} 
        \\ \hline
        
        {Internet \newline Connectivity} & {Autonomous operation (e.g., WiFi)} & {Autonomous operation or intermittent Internet connectivity} & {Autonomous operation or mobile Internet} & {Autonomous operation or mobile Internet} 
        \\ \hline
        
        {Service \newline Coverage} & {Local} & {Less global} & {Less global} & {Global} 
        \\ \hline
        
        {Latency} & {Low} & {Low} & {Low} & {Medium} 
        \\ \hline
        
        {User \newline Proximity} & {Low} & {Relatively low} & {Low} & {Various, from low to long} 
        \\ \hline
        
        {Architecture} & {Localized} & {Distributed/hierarchical} & {Localized/hierarchical} & {Localized/hierarchical/on-demand} 
        \\ \hline
        
        {Storage \newline Capacity} & \multicolumn{2}{c|}{Dependent on deployment scenarios \cite{mukherjee2018survey,pham2020sum}} & {High} & {High} 
        \\ \hline
        
        {Computational Power} & \multicolumn{2}{c|}{Dependent on deployment scenarios \cite{mukherjee2018survey,pham2020sum}} & {High} & {High} 
        \\ \hline
        
        {Power \newline Consumption} & {Low \cite{yousefpour2019all}} & {Low \cite{yousefpour2019all}} & {High \cite{yousefpour2019all}} & {Low} 
        \\ \hline
        
        {Availability} & \multicolumn{3}{c|}{High \cite{roman2018mobile}} & {Very high} 
        \\ \hline
        
        {Scalability} & {Low} & \multicolumn{2}{c|}{High \cite{roman2018mobile}} & {Very high} 
        \\ \hline
        
        {Mobility} & {Low \cite{roman2018mobile}} & \multicolumn{2}{c|}{High \cite{roman2018mobile}} & {Very high} 
        \\ \hline
        
        {Security} & {Medium} & \multicolumn{3}{c|}{High}
        \\ \hline
        
        {Context awareness} & {Medium} & {High} & {High} & {High} 
        \\ \hline
        
        {Local \newline Awareness} & \multicolumn{4}{c|}{High \cite{danaweera2021Survey}}
        \\ \hline
        
        {Standard \newline Organizations} & {NIST} & {OpenFog Consortium} & {ETSI, 3GPP, ITU-T} & {-} 
        \\ \hline
        
        {Virtualization support} & {VM} & \multicolumn{3}{c|}{VM, NFV, network slicing, and other virtualization technologies} 
        \\ \hline
        
        {Operation Mode} & {Standalone only} & {Cloud-connected support} & {Both cloud-connected and standalone operation} & {Both cloud-connected and standalone operation} 
        \\ \hline
    \end{tabular}
    \label{Tab:EC_comparison}
\end{table*}

After analyzing these features, we draw a comparison between aerial computing and other edge computing paradigms, as presented in Table~\ref{Tab:EC_comparison}. Based on the above explanations, aerial computing can support other computing concepts, such as in-network computing and cloud-native computing. However, resource and operation constraints of aerial computing limit its capability to enable high-performance services in in-network computing and scalable applications in cloud-native computing. Therefore, the integration of aerial computing with edge computing, in-network computing, and AI techniques is necessary to provide advanced services and applications in the future. For example, network elements in aerial computing, as edge devices, can partially perform computation tasks before pushing them into the network elements (e.g., switches and routers) for further processing. The integration shows great potential to respond to heterogeneous network scenarios and application requirements in 6G, but several challenges and limitations should be addressed \cite{zeng2021guest}.

\subsection{Network Design}
\label{sec:Design_Design}
This part explains the important models for designing and optimizing an aerial computing system.
\subsubsection{Computation Model}
The first component of aerial computing involves computing models with the following fundamental aspects:
\begin{itemize}
    \item computation offloading decision, 
    
    \item deployment locations of edge servers, 
    
    \item collaborative and hierarchical computation.
\end{itemize}

\textit{Computation offloading decision}: A computation task $I$ can be modeled by a tuple $I=(C,D,T)$, where $C$ is the computation workload (CPU cycles per bit), $D$ is the task size (bits), and $T$ is the completion time deadline (seconds). Fundamentally, an offloading decision results in either local computing at the devices, full execution at the edge server, or partial offloading at both. In the case of partial offloading, the task size $D$ is divided into smaller parts for local execution and remote execution at the edge servers and/or nearby devices. 

\textit{Deployment location}: Computation in aerial computing may have the following deployment location possibilities: 1) local devices, 2) terrestrial edge servers, 3) LAC servers, 4) HAC servers, 5) LEO satellites, and 6) centralized clouds. Each deployment location of the edge servers has both advantages and disadvantages. For example, the first scheme (i.e., local computing at the end devices) does not incur any transmission latency, but poor computing resources render it unsuitable for many emerging applications and new services, such as extended reality (XR) and smart cities. Further, computation and content caching at the satellite computing platforms (e.g., LEO satellites) is highly beneficial to applications with users in remote areas (e.g., mountains and islands) but might result in a higher latency and hardware cost compared to those of terrestrial computing and LAC/HAC platforms \cite{zhang2019satellite, xie2020satellite}.

\begin{figure}[t]
\centering
\includegraphics[width=1.0\linewidth]{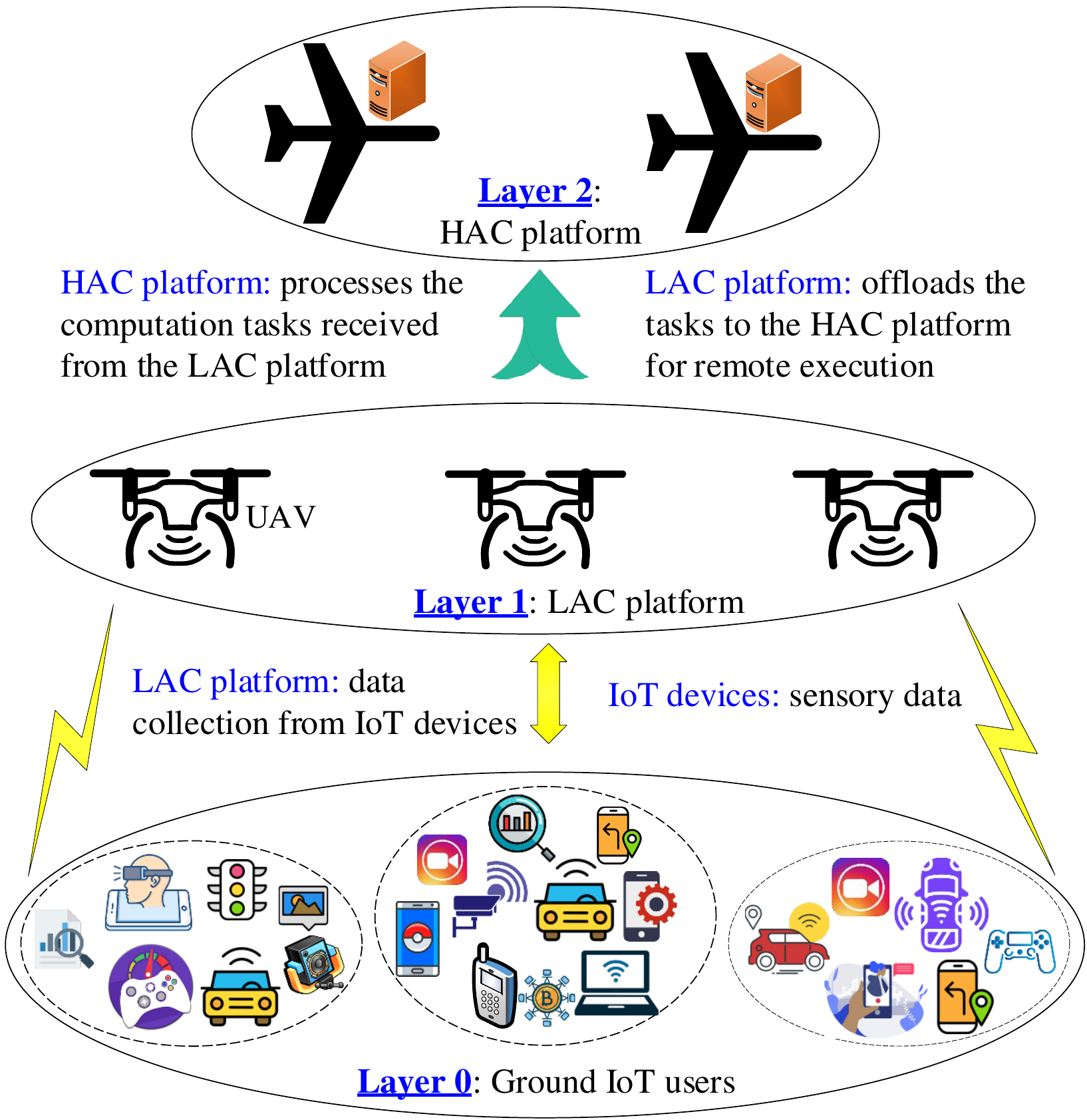}
\caption{Collaborative computation between LAC and HAC platforms.
} 
\label{Fig:2_tier_computing}
\end{figure}

\textit{Collaborative and hierarchical computation}: However, reliance on only an aerial computing node/platform (i.e., LAC or HAC only) to perform emerging compute-intensive applications is highly impractical. This is due to the onboard energy limitations of LAC platforms, intermittent harvested energy of HAC platforms, and hardware cost of satellite computing platforms for big data applications. The synergy between LAC and HAC platforms in aerial computing systems is discussed in \cite{liu2019minimization}; herein, it is illustrated in Fig.~\ref{Fig:2_tier_computing}. In such a scenario, the more powerful HAC platform and fixed-wing LAPs can assist the long-term operation and computation  tasks of rotary-wing LAPs, which are mainly responsible for data collection from ground IoT users \cite{liu2019minimization, xie2021exploiting}.

\subsubsection{Communication Model} The communication model plays an important role in aerial computing systems for guaranteeing efficient communications among computing platforms (e.g., between IoT devices and LAC servers). Several communication models have been proposed for this purpose.
In the literature, orthogonal multiple access (OMA), NOMA, and MIMO have been typically adopted for the achievable rate formula. When OMA (e.g., time division multiple access (TDMA), frequency division multiple access (FDMA), and orthogonal frequency-division multiple access (OFDMA)) are widely used for ease interference management. In addition, there are studies on hybrid OMA-NOMA computing systems \cite{liu2021latency}. 
They demonstrated that hybrid NOMA systems are highly advantageous compared to conventional OMA systems. Thus, the use of such hybrid OMA-NOMA approaches in aerial computing systems is a promising approach.

\subsubsection{Energy Consumption Model}
Energy efficiency is an important design requirement for aerial computing systems. Energy is consumed in several aspects, including local computation, communication, remote computing, and operation. 
The other part is the energy consumed by the aerial components (i.e., aerial users and servers), which is important for designing sustainable operations of aerial computing systems, particularly LAC platforms with limited onboard batteries. However, various factors, such as flying modes, flying speed, payload, and external conditions (e.g., weather and wind speed) can affect the energy model of aerial components. In an effort to incorporate this, the authors of \cite{dorling2017vehicle} modeled the power consumption of multirotor helicopters in hover and vertical moving modes. The power consumed by the UAV to remain afloat in the horizontal moving mode is modeled in \cite{austin2011unmanned}.
These models have been studied to improve the energy and power efficiency of aerial computing \cite{zhu2021uav}. Moreover, the possibility of aerial computing in terms of energy and power consumption can be realized by integrating aerial computing with emerging technologies in 6G, as discussed in Section~\ref{sec:Technologies}.

\subsubsection{Latency Model}
A benefit of aerial computing is the incorporation of LEO satellite computing platforms, which provide a global coverage but have lower latency compared to geostationary Earth orbit (GEO) and medium
Earth orbit (MEO) platforms. As reported in \cite{su2019broadband}, the OneWeb LEO satellite system can provide Internet services for 1 billion users with an average latency of 30 ms. The Starlink system promises a latency between 25 and 35 ms, while the Telesat system aims to achieve latency from 30 to 50 ms, and the average latency of the SpaceX system is less than 20 ms. Moreover, positioned in the lowest tier of the comprehensive aerial computing architecture, LAC platforms are expected to provide much lower latency and thus are suitable for latency-critical services. Further, in the middle tier, HAC platforms aim to balance latency and computing service availability. Therefore, aerial computing can provide computing services with different latency requirements while offering desirable features, such as mobility, availability, scalability, flexibility, and simultaneity. 

Other latency sources in aerial computing include computation, queuing, and fronthaul and backhaul transmission. Latency also occurs when the server executes the computation task remotely and broadcasts the processed results to the users in the downlink. Consequently, aiming to minimize the maximum latency among users, \cite{zhou2020communication} investigated a joint problem of two-dimensional LAC placement, fronthaul and backhaul bandwidth allocation, computing resource allocation, and caching decision. Further, in \cite{zhang2019joint}, the joint optimization of task allocation, scheduling, power control, and LAC server trajectory was studied to minimize the total energy consumption. To manage the latency requirement, the LAC server must execute a minimum number of bits offloaded from each user.

\subsection{Summary}
This section presents the system architecture of aerial computing, fundamental features, and key design aspects. Complementing conventional terrestrial computing systems, the proposed aerial computing framework provides computing services to massive IoT users worldwide. Compared to other edge computing paradigms, aerial computing offers several desirable features, such as mobility, availability, scalability, flexibility, and simultaneity. Moreover, four key design aspects of aerial computing have been reviewed, namely, computation, communication, energy consumption, and latency. 
Because many emerging applications will be available in the near future, more thorough investigations are required to explore the potential advantages of aerial computing. 

\section{Enabling Technologies}
\label{sec:Technologies}
\begin{figure*}[t]
\centering
\includegraphics[width=0.975\linewidth]{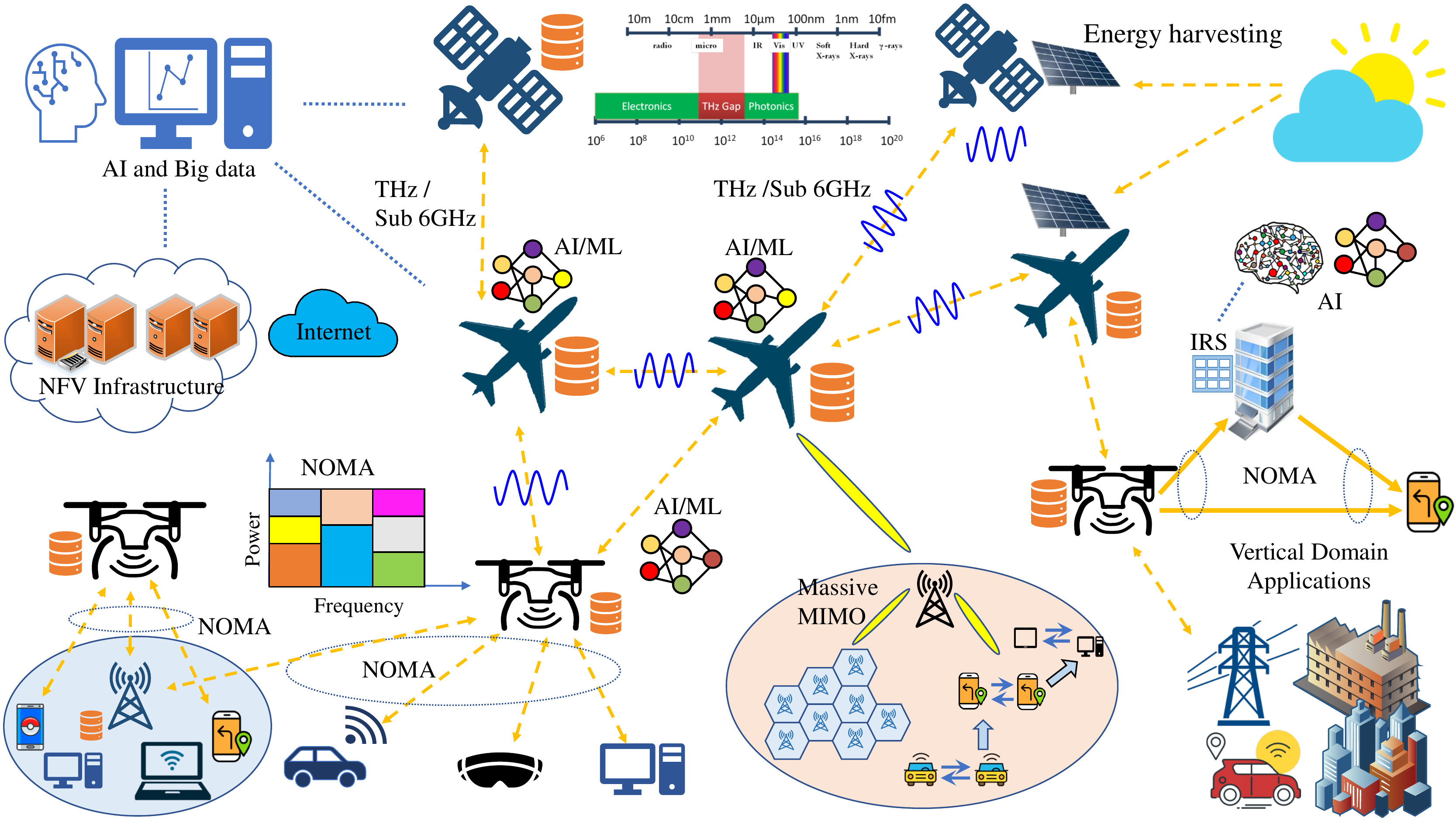}
\caption{Integration of aerial computing with enabling technologies.}
\label{Fig:EnablingTechnologies}
\end{figure*}

Enabling technologies of aerial computing are presented in this section. These technologies include network softwarization, energy refilling, edge computing, frequency spectrum, multi-access techniques, AI, and big data. To enrich the
understanding of the integration of aerial computing with enabling technologies, we illustrate the manner in which enabling technologies support aerial computing in Fig.~\ref{Fig:EnablingTechnologies}. 

\subsection{Network Softwarization}
Virtualization has enabled network operators to design, implement, and manage systems and network services with improved efficiency. NFV and SDN are two promising technologies that facilitate virtualization. NFV decouples network functions from proprietary hardware (e.g., firewalls and routers), thereby providing equivalent network functionality with general-purpose servers. SDN separates the control plane from the forwarding plane, thereby enabling network operators to configure and manage network functions in a centralized manner through the software. Other commonly used virtualization technologies include network slicing and VM. We briefly introduce the fundamentals of these technologies and thereafter discuss their application in aerial computing. 

\subsubsection{Network Function Virtualization}
The concept of NFV was first proposed in 2012, with an aim to reduce the capital and operating expenses of telecom networks, as well as to facilitate the deployment of new services. Its primary idea is to decompose the network functions from physical network equipment via virtualized network functions (VNFs). A typical NFV architecture consists of three main elements: infrastructure, management, and orchestration. The NFV infrastructure represents computing, storage, and network hardware as well as software (abstractions of physical resources) resources constituting the environment where VNFs are deployed. In addition, a VNF is an implementation of a network function, for example, a firewall, deployed on virtual resources such as VMs. Furthermore, NFV-Management and Orchestration (MANO) is the management and orchestration framework needed to provide the VNFs. 

NFV provides many opportunities for network service provisioning, e.g., independent deployment and maintenance of software and hardware, flexible and agile service deployment, and dynamic resource allocation. Therefore, NFV can benefit aerial computing. For instance, NFV can allocate additional resources from another LAC/HAC/satellite computing platform to alleviate computing congestion in one platform. 

\subsubsection{Software-Defined Networking}

SDN is an emerging network architecture that intelligently controls the network by decoupling the control plane and efficiently managing the network via programming operations \cite{wu2021ledge}. Driven by its key features, including separated control plane and data plane, centralized controller, open interfaces, and programmable operations, an SDN can intelligently operate the network at a low operating cost. Thus, it can support intelligent applications through simplified hardware, software, and management \cite{taleb2017multi}. 
Despite these advantages, SDN has several key challenges in the 6G network: 1) efficient and intelligent maintenance of the dynamic network topology \cite{SDN17}, 2) intelligent network management and orchestration via AI/ML, and 3) traffic engineering with heterogeneous QoS requirements \cite{SDN2019}.

In aerial computing, wireless networks support heterogeneous IoT devices/users, LAC/HAC platforms, 6G technologies, and configuration interfaces. SDN can provide logically centralized control by abstracting the underlying network infrastructure. For instance, SDN controllers can handle aerial-computing-related VNFs and VMs as another type of resource that can be dynamically managed. Therefore, SDN can provide an efficient solution for managing aerial computing systems and services.

\subsubsection{Network Slicing}
Network slicing is a virtualization technology that allows multiple logical networks (slices) to run on a unified physical network infrastructure \cite{nkenyereye2021virtual}. Each logical network is independently configured to satisfy the required network characteristics, such as bandwidth, delay, and capacity, to provide diverse services of expected scenarios. Each logical network also contains computing and storage resources that are capable of realizing specific network functions via NFV or service function chains. In contrast, VM can provide fine-grained control to instantiate and terminate tasks and processes at any time without affecting the hardware on aerial platforms. 
Each VM shares computing, storage, memory, and network resources from aerial computing systems, while its operation is entirely isolated from that of the host and guest VMs. Further, AI/ML can be employed to intelligently allocate virtual resources of VMs in aerial computing.

In aerial computing, network slicing can be utilized to slice the entire network into individual networks considering computing services and IoT services, which can be optimized by specific requirements and services. However, an AI-enabled intelligent management and orchestration framework should be designed to efficiently support network slicing in 6G networks.  
\subsection{Energy Refilling}
\subsubsection{Energy Harvesting}
Energy harvesting is a novel approach to utilize renewable energy sources. For example, solar energy can be converted into electric energy by deploying photovoltaic (PV) cells on UAVs, and the electric energy can be stored in rechargeable batteries \cite{EHMagazine2019}. However, natural energy sources are highly dependent on climate variability. For instance, solar power harvesting can be severely affected by weather conditions.

Some non-electromagnetic field-based charging technologies have been used, including gust soaring, PV arrays, laser beaming, and battery dumping \cite{WChargingAccess18}. Dynamic soaring involves harvesting energy from wind and airflow by adjusting the trajectory of UAVs to prolong their flying duration \cite{chittoor2021review}. For PV arrays, solar irradiation energy is harvested by PV cells to power a drone or recharge its battery, which is used for the operation during the night. This technology can be applied to fixed-wing drones because a certain amount of space is needed for PV panels. In laser-beaming technology, a laser fed by an external source of energy produces a concentrated and streamlined beam of light with a certain frequency to the specific PV cells on a drone. The PV cell can then convert the laser beam into energy to recharge the battery of the drone.

\subsubsection{Wireless Power Transfer}
Another popular application of energy harvesting is that energy is usually harvested from ambient radio frequency electromagnetic signals. In simultaneous wireless and information power transfer (SWIPT), energy can be harvested, usually partially, from the transmitted signals via power switching or time switching techniques. A strong user with a high channel gain can forward data to a weak user with a low channel gain by using the harvested energy from the transmitted signals. Wireless power communication networks (WPCNs) are used to harvest energy from radio transmission, wherein sequential energy is transferred wirelessly to the intended users, which supports dedicated wireless charging \cite{NOMAMag2019}. 
Aerial components at different computing platforms (e.g., UAVs at the LAC platform and balloons at the HAC platform) have been proposed to be integrated with WPCN. This flexible implementation allows aerial components to be charged or to charge a set of ground users, thus providing sustainable solutions for aerial computing systems \cite{pham2021uav, xie2021uav}. Besides, aerial components can be equipped with solar-powered cells and large energy storage to store energy harvested from solar powers in the environment.

Wireless power transfer (WPT) has been proven to provide power for UAVs wirelessly without landing for energy refueling \cite{UAVWPS2020}. Moreover, the coordination of LAC servers at the LAC platform and edge servers on different platforms (i.e., LAC, HAC, and satellite) is a promising approach to better provide computing services over large-scale areas \cite{xie2021uav}. Since WPT from the air is quite limited due to the severe path loss, many factors (e.g., antenna size) need to be considered for practical implementation. To further support low-power IoT/mobile devices, emerging technologies, such as backscatter communications and IRS, can be deployed. For example, \cite{nguyen2021backscatter} considered IoT devices as backscatter-assisted users, which harvest energy from the ground gateway to perform computation offloading. Similarly, aerial components can assist low-power users in performing necessary data transmissions and computation tasks.

\subsection{Frequency Spectrum}
\subsubsection{Sub-6 GHz and mmWave}
To meet the extremely high data rate requirements of 5G systems, high-frequency spectrum bands are exploited. In particular, millimeter waves (mmWave) can provide gigabit-per-second data rates owing to the radio frequency spectrum in the range of 30--300 GHz, whereas the sub-6 GHz range can provide large coverage with high cost efficiency with a spectrum band below 6 GHz \cite{chamitha2021frontiers6g}. 
Driven by the increase in IoT devices, such as XR services encompassing AR, mixed reality, and VR, some studies proposed the integration of mmWave and Sub-6 GHz to balance the high data rate and large coverage in next-generation wireless communication systems. Specifically, mmWave can be used in tiny cells for mobile and fixed access, whereas sub-6 GHz can be used in small cells. For example, the use of the mmWave band in aerial computing was considered in \cite{jiao2021intelligent}, wherein a hierarchical architecture of a satellite computing server, two terrestrial MEC servers, and a set of vehicles was considered.

\subsubsection{THz}
Terahertz (THz) band, in the frequency range of 0.1--10 THz, has been explored in the next-generation wireless communication systems to achieve the terabits per second data rate with low latency. Potential use cases for THz communications include close proximity communication, extremely high data rates for indoor communications, and wireless backhauling and fronthauling technologies. Thus, THz can be utilized in LoS and high data transmission among aerial components in aerial computing systems \cite{THz2021Chen}.

\subsubsection{Other High-Frequency Bands}
At WRC-12, the ITU identified the C-band for the use of control and non-payload communication. The C-band refers to a frequency range of 4--8 GHz and is used primarily for satellite communications. Moreover, the K-band in the frequency range of 14--27 GHz is typically used for short-range communications owing to the high atmospheric attenuation.
The K-band can be used in satellite communications, astronomical observations, and radar. Radars in this frequency range provide a short range and high throughput.

\subsection{Artificial Intelligence and Big Data Analytics}
AI technology aims to train machines to perform human tasks. It has been applied to various areas, such as image recognition, robotic vehicles, machine translation, and game AI. Further, ML, which is a promising subset of the AI technique to learn from the data and impart intelligence to existing systems, has been employed to render wireless communication and networks highly efficient and adaptable. Intelligent wireless network can provide efficient support for aerial computing. The advantage of applying ML in wireless network operation is that it enables the network to monitor, learn, and predict various communication-related parameters, such as wireless channels, traffic patterns, user context, and device locations. In general, ML algorithms include supervised learning, unsupervised learning, reinforcement learning (RL), deep learning (DL), deep reinforcement learning (DRL), and federated learning (FL) \cite{MLsurvey19}.

\subsubsection{Supervised Learning/Unsupervised Learning}
In supervised learning algorithms, both the input and desired outputs of the used datasets are available. However, supervised learning algorithms can only be employed in scenarios with sufficient labeled data, for example, classification. In contrast, in unsupervised learning, the dataset used for training does not include labeled output or target values. The purpose of unsupervised learning is to extract key features of the data for better prediction. Further, unsupervised learning algorithms can be applied to scenarios, such as clustering based on the available data. In aerial computing, supervised learning/unsupervised learning can be used for user clustering/grouping to provide better service.

\subsubsection{Deep Learning} 
DL algorithms have been developed to deal with complex input--output mappings. DL consists of multiple layers for feature extraction and transformation. 
DL can be used for in-depth analysis in a complex scenario with massive data and to realize different control schemes for different protocol layers. For example, in \cite{bai2021lscidmr}, DL was used to classify cloud images captured by satellite clouds in aerial computing systems.

\subsubsection{Reinforcement Learning} 
The main idea of RL is to train the agent to generate actions according to the current environment. In RL, the problems are solved by employing a sequence of actions that use the trial-and-error rule. RL algorithms have been extensively used in wireless network optimization to obtain the optimal policy, for example, user grouping decisions or actions. DRL is a DL framework developed based on RL; it relies on updated samples in practice instead of the ideal transition probability in theory. DRL involves learning from the feedback that evaluates the actions taken rather than learning from the correct actions. DRL algorithms have been applied to wireless networks for multiple aspects, including mobile networking, resource allocation, schedular design, and routing \cite{AI2021She}. Furthermore, it can promptly make a decision under dynamically changing network conditions, such as channel state information \cite{qian2021noma}.

\begin{figure*}[t]
\centering
\includegraphics[width=1.0\linewidth]{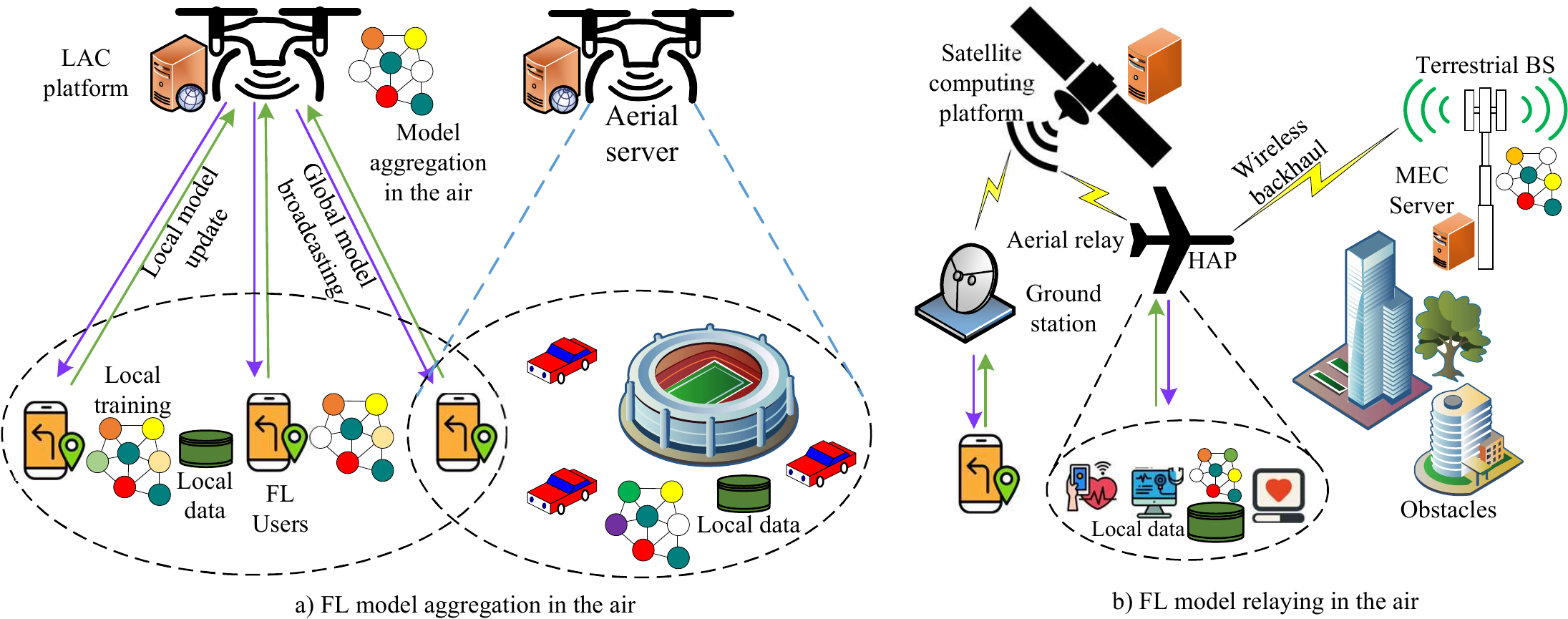}
\caption{Applications of aerial computing for FL-enabled wireless systems \cite{pham2022aerial}.}
\label{Fig:FL_aerial_computing}
\end{figure*}

\subsubsection{Federated Learning}
FL is an AI approach that enables users to collaboratively learn a shared model with data maintained on their own devices \cite{FL21}. In contrast to a standard ML algorithm that requires a centralized training dataset in the data center, FL allows devices to train a learning model locally and transmit the training parameters instead. Thus, it addresses concerns, such as user privacy and limited data transmission resources (e.g., bandwidth) \cite{FL21, pham2021uav}. A potential application of aerial computing for FL-enabled wireless networks is aerial aggregation \cite{pham2022energy}; that is, a computing server (e.g., LAC, HAC, and satellite computing) in aerial computing acts as the aerial server and performs model aggregation in the air, as shown in Fig.~\ref{Fig:FL_aerial_computing}(a)]. Moreover, as shown in Fig.~\ref{Fig:FL_aerial_computing}(b), aerial computing can extend the learning coverage of terrestrial FL-enabled networks, particularly when certain obstacles prevent model updating and broadcasting between FL users and the aggregation server. 
Further, aerial components (e.g., LAC, HAC, and satellites) can be FL users, which train local AI models and share updates with the central cloud \cite{fadlullah2021smart}.

\subsubsection{Big Data}
The features of big data have been commonly described by five “V”s: volume, variety, velocity, veracity, and value. Certain attempts have been undertaken to apply big data analytics to wireless communications \cite{Bigdata14}. The increasing complexity of networks and complicated wireless traffic patterns make big data analytics appealing. ML algorithms coupled with edge computing can be used to process big data in 6G \cite{Bigdata20}. MEC has become the primary computing method for big data analysis in heterogeneous 6G.

AI and big data, combined with other enabling technologies mentioned in this section, have been widely introduced to aerial computing in 6G owing to the advantages of intelligent management and automatic self-improvement. In particular, AI/ML can be used in intelligent UAV/drone trajectory design, virtual resource management, data processing, task computation, and channel estimation. In addition, AI and big data can be used in NFV and SDN-based networks to achieve intelligent network management and optimization \cite{AI2018Guo}. RL approaches have also been conducted to improve resource utilization and service provision \cite{AI2017Kim}. 
Further, advantages of the other ML techniques, such as meta learning and transfer learning, can be exploited. For example, the benefits of transfer learning with knowledge preprocessing in LAC systems have been demonstrated in \cite{zhang2021transfer}. However, major challenges should be answered to further improve the effectiveness of transfer learning in heterogeneous aerial computing systems. More specifically, we need to determine the source agent and transferred knowledge, and optimize scheduling and resources.

Certain challenges need to be addressed to realize aerial computing in 6G, including data security/privacy, learning efficiency, communication cost, and the tradeoff between learning accuracy and convergence. AI/ML is recognized as the most promising solution for providing intelligent wireless communications and signal processing computation task control for increasingly complex and heterogeneous networks.

\begin{table*}[ht!]
    \renewcommand{\arraystretch}{1.20}
	\caption{Summary of enabling technologies of aerial computing.}
	\label{Table:Enabling Technologies}
	\centering
	\begin{tabular}{|p{1.9cm}|p{1.9cm}|p{5cm}|p{7cm}|}
		\hline 
		\textbf{Category} & \textbf{Enabler} & \textbf{Advantages}&\textbf{Key Challenges}\\ 
		\hline	\multirow{4}{2.2cm}{Network Softwarization \cite{wu2021ledge, SezerSDN, SDN17, SDN2019, AfolabiSurvey, BenzaidAIDriven}}
		& NFV & High flexibility and low operational cost & Service heterogeneity  \\ 
		\cline{2-4} 
		& VM
		& Sufficient computing resources 
		& Intelligent management of virtual resources \\
		\cline{2-4} 
		& Network slicing
		& Flexibility, resource efficiency and Security 
		& Intelligent management and orchestration framework design\\
		\cline{2-4} 
		&SDN
		& High flexibility and low operational cost & SDN controller placement\\
		\hline
		\multirow{2}{2.2cm}{Energy Refilling \cite{EHMagazine2019, WChargingAccess18, RichardsonUAV15, NOMAMag2019, xie2019throughput, xie2021uav}} 
		& Energy harvesting 
		& Ambient natural energy source 
		& Highly dependent on weather conditions \\ 
		\cline{2-4} 
		& WPT (SWIPT and WPCNs)
		& Harvest energy from radio transmission & Trade-off between harvested energy and consumed power and joint resource allocation\\
		\hline
		\multirow{3}{2.2cm}{Frequency Spectrum \cite{chamitha2021frontiers6g, THz2021Chen}} & Sub-6 GHz & High bandwidth and high data rate & Severe attenuation and blockage and low range \\ 
		\cline{2-4} 
		& THz
		& High bandwidth and high data rate  & Severe attenuation and blockage and low range\\
		\cline{2-4} 
		& C-Band  K-Band
		& Satellite communications and high throughput & Atmospheric attenuation\\
		\hline
		\multirow{3}{2.2cm}{Multiple Access Techniques \cite{DingMag17, MassiveMIMO16, MIMOMag16, MIMONOMA19, IRS2020}} & NOMA & High spectral efficiency and massive connectivity & Decoding complexity for multiple users \\ 
		\cline{2-4} 
		& Massive MIMO
		& High statistical multiplexing gain and high spectral efficiency 
		& High implementation cost, intelligent environment adaption and channel prediction\\
		\cline{2-4} 
		& IRS
		& High channel gain and low implementation costs 
		& Intelligent optimization of IRS\\
		\hline
			\multirow{4}{2.2cm}{AI/Big Data Analytics \cite{MLsurvey19, AI2021She, qian2021noma, FL21, Bigdata14, Bigdata20, AI2018Guo, AI2017Kim}} & ML 
			& Automatic features, predictions 
			& Computational complexity\ \\ 
		\cline{2-4} 
		& DL
		& Automatic features, predictions 
		& Computational complexity\\
				\cline{2-4} 
		& RL
		& Generate optimal decisions in a dynamic environment and require the Markovian model and high computation 
		& High dimensionality of the states and action spaces \\
		\cline{2-4} 
		& FL
		& Protects user data privacy 
		& Communication cost and tradeoff between learning accuracy and convergence \\
		\cline{2-4} 
		& Big data \newline analytics
		& Prediction of user preference distribution 
		& Feature extraction and data modeling\\
		\hline			
	\end{tabular}
\end{table*}

\subsection{Other Wireless Techniques}
\subsubsection{Non-orthogonal Multiple Access} 
NOMA is considered a promising technology in 5G and beyond owing to its high spectrum efficiency and massive connectivity. It is broadly classified into two categories: power-domain NOMA and code-domain NOMA. Power-domain NOMA uses the power domain for multiple access, whereas code-domain NOMA exploits sparse code for multiple access. NOMA can support more users than the number of available subcarriers, thereby improving wireless communication performance with multiple features, including massive connectivity, low latency, high spectral efficiency, and high energy efficiency.
\begin{itemize}
	\item First, as NOMA enables multiple users to be multiplexed on the same channel simultaneously, it is suitable for a scenario with a large number of connections (e.g., IoT applications). Therefore, NOMA can be applied in aerial computing to support multiple devices/users to offload their tasks to aerial servers, as shown in Fig. \ref{Fig:EnablingTechnologies}.
    \item Second, compared with OMA where users must wait until a resource block is available to transmit or receive data, NOMA can provide grant-free transmission with flexible scheduling. Therefore, the offloading/downloading delay of the devices/users in aerial computing can be significantly reduced.
    \item Third, NOMA can achieve higher spectral efficiency and user fairness than OMA \cite{DingMag17}. Further, NOMA users can utilize the entire frequency bands for transmission, whereas OMA users can only use a fraction of the entire spectrum for communication. In aerial computing, high spectrum efficiency can also be achieved in the communications between LAC/HAC platforms when using a high-frequency spectrum \cite{jiao2021intelligent}.
\end{itemize}

\subsubsection{Massive MIMO}
The concept of massive MIMO has been proposed to drastically increase data rates, spectral/energy efficiency, and coverage of wireless networks by aggressive spatial multiplexing. In massive MIMO, a BS equipped with a few hundred antenna arrays serves tens of users simultaneously. It is considered a promising technology in 6G owing to the following advantages \cite{MassiveMIMO16}.
\begin{itemize}
	\item \textit{Multiplexing gain:} Aggressive spatial multiplexing can increase the data rate dramatically in the spatial domain. Further, the offloading delay of the computation task is significantly reduced for remote aerial computing.
	\item \textit{Energy/spectral efficiency:} The energy efficiency can be improved via massive MIMO as it is inversely proportional to the number of antennas at the BS. 
	\item \textit{Increased robustness:} Due to a large number of antennas, the propagation channel provides additional diversity gains, which also increases the link reliability. Thus, MIMO can provide reliable links in aerial computing, where the uncorrelated noise and intra-cell interference can vanish with an increasing number of antennas.
\end{itemize}

Various multiple access technologies, such as OMA and NOMA, can be combined with massive MIMO to further improve the communication performance \cite{zeng2020massive}. In massive MIMO-OMA systems, the maximum ratio combing and zero-forcing can be utilized to achieve a high spectral efficiency in underloaded systems. However, some studies proposed massive MIMO-NOMA to support a large number of users; that is, the overloaded system, and to meet the massive connectivity requirement of 6G \cite{MIMONOMA19}. However, there exist challenges for the realization of massive MIMO, which are the 1) deployment of extremely large antenna arrays, 2) adaptation to the intelligent wireless environment, and 3) limitation in channel prediction. Equipped with full-dimensional large arrays, ground BSs can apply adaptive fine-grained 3D beamforming to mitigate the strong interference between high-altitude UAVs and low-altitude terrestrial users.

\subsubsection{Intelligent Reflecting Surface} 
IRS has been proposed as a promising technology in 6G to alter the wireless channel by intelligently adjusting the amplitude and/or phase shift of each element of the IRS \cite{IRS2020}. Extensive deployment of IRSs in the wireless network and smart coordination of the reflections can result in the flexible reconfiguration of the wireless channels between transmitters and receivers to expand the communication coverage and improve the wireless communication capacity and reliability.
IRS has various advantages.
\begin{itemize}
	\item \textit{Low cost}: The reflecting elements of IRS are low-cost printed dipoles, which only passively reflect the signal without transmitting the signal.
	\item \textit{Full-duplex mode}: Different from the traditional relay system with half-duplex relay, IRS can work in a full-duplex mode.
	\item \textit{Auxiliary device}: IRS has great flexibility and compatibility to be integrated into the existing wireless networks such as WiFi or cellular.
\end{itemize}
Thus, IRS can be massively deployed in wireless networks and combined with other promising technologies, such as NOMA \cite{FangIRS2020}, massive MIMO, and UAVs. However, IRS is still in its nascent stages, and certain potential challenges need to be addressed: 1) The deployment of IRS, especially for air computing, the location IRS can significantly affect the channel condition for data transmission;  2) The reflection of IRS is sensitive to the angle of arrival, which affects the channel model of the IRS in practice \cite{IRS2020}.

IRS can be applied in various scenarios to improve the communication performance \cite{THz2021Chen}. The most important task for aerial computing is to prolong the battery life of UAVs, LAC servers, and IoT devices. Energy harvesting or WPT techniques can be implemented using IRS to overcome the high-power loss over long distances. For example, a highly efficient energy charging zone can be created by deploying IRSs in the proximity of devices or UAVs. In addition, if IRSs are deployed on UAVs, the UAVs can flexibly establish strong LoS links with the ground nodes to improve the communication quality for aerial computation task offloading and downloading.

\subsection{Summary and Discussion}
In this section, we have presented a set of enabling technologies of aerial computing. Table~\ref{Table:Enabling Technologies} presents a summary of the enabling technologies and their advantages and challenges. These technologies can be flexibly implemented in future 6G networks to efficiently provide fast computing services, better mobility, and higher scalability and availability. Further, these technologies can facilitate the process of addressing challenges, such as intelligent communication and computation resource allocation design, joint optimization of UAV trajectory and placement design, user association and grouping design, and secure computing and communications. However, certain challenges need to be addressed, as highlighted below.
\begin{itemize}
	\item \textit{Intelligent Control:} The cooperation of enabling technologies should allow for intelligence in aerial computing. The joint intelligent optimization of computing resources (LAC, HAC, and satellite computing platforms), communication link design with different multi-access techniques, and network control/scheduling can be investigated in future studies.
	
	\item \textit{Secure Control:} For aerial computing, the user data need to be transmitted and shared with the aerial server, which leads to security concerns and data transmission overhead. Therefore, FL and blockchain can be exploited to protect data privacy and improve security by transmitting the training parameter to the server for global aggregation \cite{lim2020federated, nguyen2021federated}.
	
	\item \textit{Aerial Intelligence:}
	AC, HAC, and satellite computing platforms can enable aerial intelligence by collecting surrounding data and then executing onboard intelligent algorithms. The aims of intelligent algorithms include autonomous collision avoidance, adaptive flight gesture adjustment, and trajectory optimization for data collection. Therefore, the design of intelligent algorithms can efficiently support aerial computing in terms of energy savings, large coverage, delay minimization, etc.
\end{itemize}

\section{Vertical Domain Applications}
\label{sec:Applications}
Aerial computing is expected to serve different vertical domain applications that require service provision from different perspectives. In this section, we explore the impact of aerial computing on relevant domains: smart cities, smart vehicles, manufacturing, and smart grids. 

\subsection{Smart City}
The IoT involves all devices that are connected to the Internet on Earth. Due to the potential of these devices in improving the quality of human lives, they are sensitive to latency, storage, bandwidth, and security. With the increasing popularity of the IoT, cities are becoming smarter, and the concept of smart cities is gaining traction and a new dimension. Aerial computing can aid in better understanding the potential of the IoT in developing new strategies for smart cities, thus reducing costs and improving safety. Furthermore, new endeavors, such as 5Gcity~\cite{Colman2019} and SynchroniCity~\cite{Cirillo2020}, can help realize the idea of a future city. According to~\cite{Colman2019}, there are four possible major themes in smart cities, which are data communication and processing, MEC support, urban planning and management, and surveillance and security. In Fig.~\ref{Fig:smart_city}, we provide an example of a smart city through the use of aerial computing technologies.

\subsubsection{Data Communication and Processing}
\begin{figure}[t]
\centering
\includegraphics[width=1.0\linewidth]{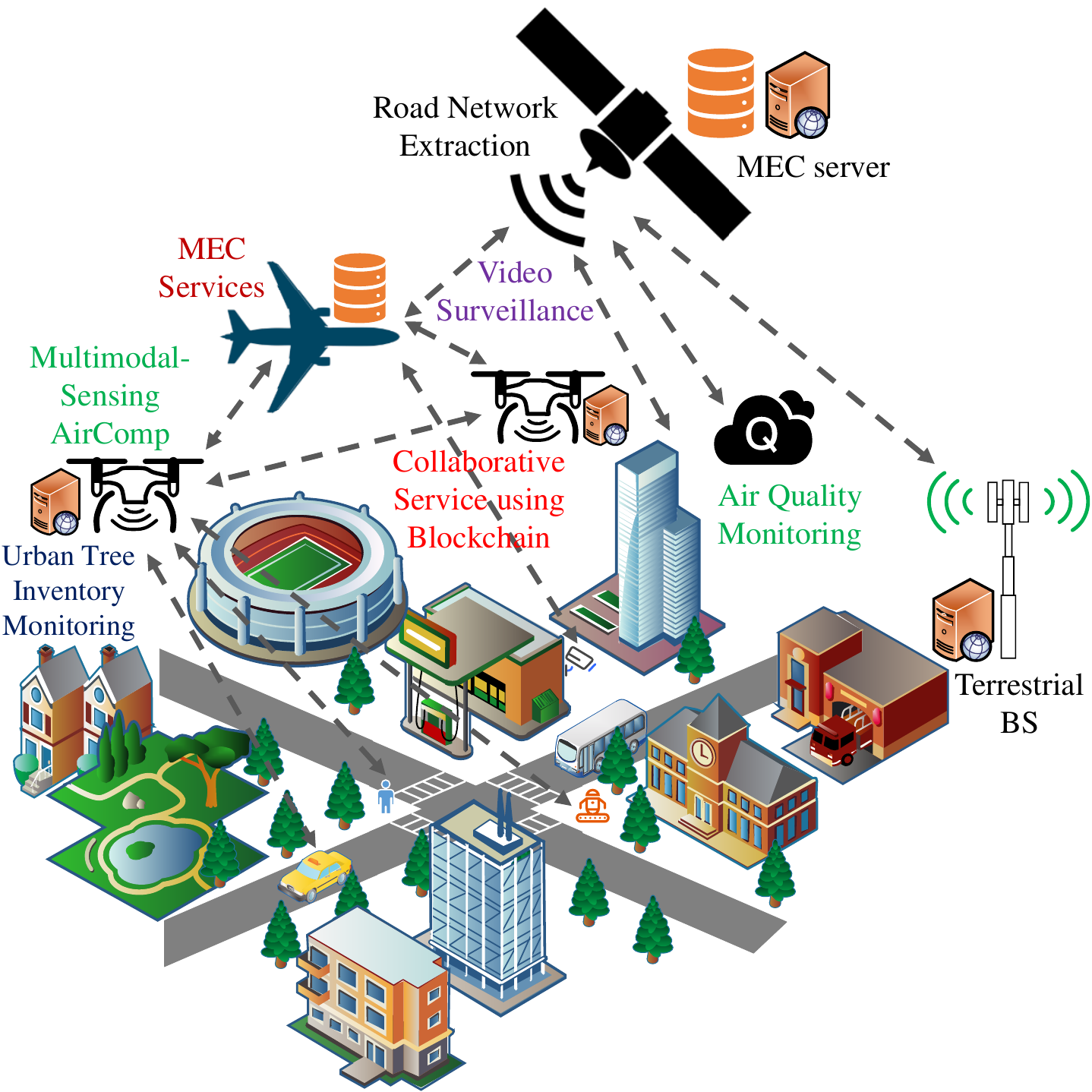}
\caption{Illustration of a smart city using aerial computing.} 
\label{Fig:smart_city}
\end{figure}

With embedded multiple sensors, LACs, HACs, and satellite computing platforms have great potential for sensing data in IoT environments and providing social services to smart cities. In \cite{YLiu2021}, the authors proposed a joint scheme of 3D placement, computation, and communication resource allocation for multiple HACs in an uplink IoT network, where the task distribution among HACs and communication resources were considered. The solution scheme was developed based on the K-means method, a Hungarian-based algorithm, and an iterative method. The simulation results illustrate that the total transmission power of the IoT nodes is significantly minimized through the proposed algorithm. In \cite{GZhu2019}, a technique referred to as AirComp was developed to receive data from multiple sensors simultaneously by utilizing the superposition property of multi-access channels, as shown in Fig.~\ref{Fig:smart_city}. A near-optimal equalizer was derived using differential geometry to facilitate multimodal sensing technology in the proposed technique. Finally, an efficient channel feedback mechanism was designed to facilitate the acquisition of the entire channel information from many sensors simultaneously. In \cite{CHLiu2019}, a novel and efficient control algorithm for LACs was proposed to manage sensing and their movement using DRL techniques. The proposed method extracts features using the convolutional neural network (CNN) and infers decisions following the multi-agent deep deterministic policy gradient (DDPG) method.  However, swarms of LACs and HACs may face difficulties owing to their limited storage and computation capacities during crowd sensing. To resolve this issue, in \cite{WChen2019}, edge/cloud computing technologies were introduced to the swarm for enhancing their QoS. Based on a case study of latency-critical applications, their simulation results illustrated that the proposed approach could effectively improve the QoS of the swarm servicing entity. Furthermore, to enhance the poor coverage of information-centric IoT networks, the authors of \cite{TLi2019} optimized the coverage of such a network via LACs and HACs at a low cost. The proposed scheme was designed based on an improved rolling horizon strategy (IRHS). Through comprehensive experiments compared to previous studies, the authors demonstrated that the proposed scheme can improve the coverage ratio (by 21.42\%) and reduce the cost ratio (by 13.335\% to 34.32\%). 

Aiming to track a moving target using synthetic aperture radar (SAR) based on compressive sensing technology, the authors of \cite{GXu2018} considered applying the multi-channel processing technique to the SAR computation unit. Because of the joint consideration of the magnitude and phase, the proposed weight-adjusted sparse algorithm in the computing unit could improve the performance of clutter suppression. In another work~\cite{GXu2018-2}, the same authors determined a different method to track moving targets with the assistance of compensated target motion in the SAR computing unit. The motion compensation procedure in the SAR computing unit consisted of two steps: the SPECAN processing was first applied to the frequency and azimuth domain, and then the correction of the residual component was incorporated into the sparse imaging outcome. Further, the authors also overcame the challenges associated with the motion of the target and the imaging of multiple moving targets using a novel parametric sparse imaging approach and an orthogonal matching pursuit method, respectively.

\subsubsection{MEC Support}
In \cite{JWang2020}, a novel online HAC-assisted edge server management scheme was proposed to provide flexible edge computing services. Several hot-spot areas were extrapolated by geographically merging the tasks, and the corresponding HACs were dispatched to the appropriate locations for the necessary service. A simple illustration of this use case is provided in Fig.~\ref{Fig:smart_city}. Through extensive simulation with real-world data, the authors showed that the proposed mobile server dispatching scheme could serve more users with a high utilization of resources. The authors of \cite{XDiao2019-2} proposed a fairness-aware task distribution and trajectory optimization scheme in LAC systems, in which a fixed-wing LAC played the role of a flying computing server for mobile terminals. Under fairness consideration, the authors aimed to minimize the maximum energy consumption of all mobile terminals. The non-convex problem was converted into a convex one, and an iterative algorithm was proposed to solve the problem. Further, under the assumption that the communication between the LAC and mobile terminals followed the NOMA technology, the authors in \cite{XDiao2019} proposed an iterative algorithm to minimize the maximum energy consumption of all mobile terminals by controlling the trajectory, task data, and computing resources in an adaptive manner. The simulation results illustrated that the proposed algorithm could effectively reduce the highest energy consumption among mobile terminals while considering the fairness issue. The work in \cite{Ztan2020} introduced a highly diversified urban scenario, which was served by multiple LAC servers. Two non-convex programming problems were sequentially formulated to optimize the delay and energy performance of uplink and downlink communication scenarios for AR applications, which were solved by convex approximation techniques. Through experiments, the authors verified that the proposed network model could serve cellular users in a satisfactory manner. 

\subsubsection{Urban Planning and Management}
Extracting building anatomy through remote sensing technologies is crucial for urban planning and management. Considering diverse scales and appearances, assisted by HACs, the authors of \cite{XLiu2021} proposed an automatic method to extract building instances. The proposed HAC unit consisted of an improved hybrid task cascade method, which had three components for three different consecutive tasks: high-resolution representation, definition of guided anchor, and formation of focal loss. Comprehensive experiments on a real dataset revealed that the proposed method was shown to perform better compared to the conventional R-CNN method in terms of extracting building instances. As of the LAC server, in \cite{MWan2019}, a target tracking algorithm to locate LAC-captured targets, such as pedestrians and vehicles, was proposed using the sparse representation theory. Upon implementation with real data, the results prove that the proposed tracker can achieve better performance compared with existing tracking algorithms. Further, to monitor urban environments, the authors of \cite{PLi2019} proposed the use of heterogeneous aerial computing (e.g., LACs, HACs, and satellite computing) and investigated a collaborative methodology to share the workload. A simple illustration picture of this collaboration feature is provided in Fig.~\ref{Fig:smart_city}. For a particular set of degrees of freedom (e.g., a number of tasks and vehicles), an efficient algorithm was proposed to maximize the overall weight of completed tasks. Through simulations, the authors determined that the proposed algorithm significantly outperformed the other heuristic algorithms.

With the growing trend of industrialization, air pollution has reached a significant level in urban areas. Consequently, monitoring air pollution has become a matter of great concern in recent years. To this end, the authors of \cite{ZHu2019} presented an efficient and cost-effective air quality monitoring system using LACs and HACs. The three major aspects of the system proposed in \cite{ZHu2019} were data processing, deployment, and power control. The system was deployed at Peking University and Xidian University, and about 100,000 samples have been collected since February 2018. A partial illustration of this system is provided in Fig.~\ref{Fig:smart_city}. Using a 360-degree panoramic camera in HACs, the authors of \cite{JGao2021} designed a novel air quality indicator (AQI) monitoring system, AQ360, to detect the air quality level (shown in Fig.~\ref{Fig:smart_city}). Upon solving the recognition problem of images, they optimized the placement of the corresponding LAC to achieve the optimal outcome. Through the implementation of the system under realistic scenarios, they showed that the system can perform better compared to existing studies in terms of AQI recognition error and energy consumption. Further, using microwave downlink sensor networks, the authors of \cite{Colli2020} studied a rainfall monitoring system via a satellite computing platform. The proposed method was implemented in the satellite computing unit, and it was compared with the Ligurian regional tipping-bucket rain gauge (TBRG) network at a city in Italy. The performance of these two types of monitoring systems was found to be similar, based on the rain events that occurred over the time period between January 2017 and December 2018. 

Fine-grained driving lines are crucial components of high-definition maps for the smooth operation of autonomous vehicles. To this end, the authors of \cite{LMa2019} presented a semi-automated driving line generation method utilizing a mobile laser scanning computing system, as shown in Fig.~\ref{Fig:smart_city}. Through simulations on real datasets, the authors demonstrated that the proposed method achieved an average recall, precision, and F1-score of 90.79\%, 92.94\%, and 91.85\%, respectively. Further, using the same mobile laser point clouds, the authors of \cite{LLiu2020} proposed an image-translation-based method to obtain the 3D vectors of typical road markings, as shown in Fig.~\ref{Fig:smart_city}). Another method for extracting road networks using a satellite computing platform was proposed in \cite{YZang2016}, wherein a novel aperiodic directional structure measurement (ADSM) technique was adopted. In contrast, to facilitate city planning via swarms of fully automated LACs, the authors of \cite{KKuru2021} proposed a holistic distributed framework 
equipped with various effective and efficient skills, which could be adjusted based on the requirements of the LACs and environments. Through experiments, the authors demonstrated that a swarm of LACs equipped with the proposed framework could effectively satisfy the requirements of diverse applications. The authors of \cite{YChen2019} presented an efficient framework for extrapolating urban tree inventories by using 3D point clouds acquired by a HAC-borne laser scanning system (shown in Fig.~\ref{Fig:smart_city}). The high-level processing steps of the proposed HAC framework are individual tree cluster extraction, geometric parameter estimation, and tree species classification. Through experiments, the authors demonstrated that the detection accuracy of the roadside tree was over 93\% with an average error of approximately 5\%, and the overall classification accuracy was approximately 78\%. Another application of the point clouds collected by the 
computing system is the modeling of an underground parking lot~\cite{Gong2021}. The method utilized in the HAC consists of two parts: a joint localization and mapping algorithm based on sparse point clouds and a semantic modeling algorithm. The results from comprehensive experiments indicated that the proposed algorithm achieves centimeter-level accuracy with a precision of 84.8\%. 

\subsubsection{Surveillance and Security}
The concept of a smart city is vulnerable to security and privacy issues since the collection, dissemination, management, and processing of data in aerial computing platforms can be interrupted by malicious users. Thus, evidenced by certain existing reports, we discuss the way in which aerial computing platforms (e.g., LACs, HACs, and satellites) could be the means of resolving security and privacy issues in smart cities. 

With the assistance of 3D simulation and a set of robot operating system-equipped LACs, the authors of \cite{EErtugrul2018} demonstrated an autonomous navigation system to control the physical security of smart buildings. The results from the experiments indicated that the proposed approach achieves an acceptable level of accuracy in terms of mapping the indoor environment of smart buildings. In \cite{SHu2020}, a novel framework was proposed to monitor suspicious links by utilizing the characteristics of LACs. Particularly, the system consists of a suspicious transmission link and an LAC unit for monitoring. Further, in the LAC system, the wireless resource allocation problems, including trajectory planning and energy minimization, can be solved using the popular successive convex approximation method.

Video surveillance applications are associated with several challenges in smart cities, such as scalability, integrity, and latency. In \cite{YJin2020}, a series of optimal scheduling and control algorithms were designed to address these challenges. A network with full coverage was established via an LAC cluster, and the scheduling problem was solved using the bi-objective fragile bin packing technique. Thereafter, extensive simulations with realistic parameter settings were conducted, and the effectiveness of the proposed scheme was verified in terms of many systems and video-specific performance metrics. The authors of \cite{HKim2018} introduced a surveillance model for a multi-domain IoT environment using heterogeneous smart HACs and LACs (shown in Fig.~\ref{Fig:smart_city}). A heuristic method was used to minimize the maximum movement of LACs while avoiding collisions among them. Extensive simulations were conducted considering heterogeneous scenarios with multiple HACs and LACs to verify the merits of the proposed method.

In addition to limited storage and computation power, security and privacy are two of the alarming concerns for data dissemination schemes in aerial environments. To this end, the authors of \cite{GKVerma2020} presented a low-cost computationally efficient short proxy signature (CB-PS) scheme for the LAC server, which was oblivious to secret key information in conventional ID-based schemes. To further enhance the security of such tasks, the authors of \cite{Keke2020}, presented a novel blockchain-based strategy to facilitate multi-party authentication among multiple LACs and HACs. Specifically, this strategy is useful for P2P and group communications among LACs and HACs while ensuring the efficiency of data dissemination. Through comprehensive experiments and simulations, the authors verified the merit of the proposed technique in terms of authentication among multiple LACs and HACs. Subsequently, to satisfy the connectivity, data, and service demands of an exponentially growing number of IoT devices in smart cities, the authors of \cite{aloqaily2021design} envisioned a 5G network environment supported by blockchain-enabled LACs and HACs. The objective of this network was to provide Internet connectivity to IoT devices via blockchain-enabled multiple LACs and HACs and to facilitate decentralized service delivery and routing facilities in a reliable and secure manner. 

\subsection{Smart Vehicle}

Over the last 30 years, intelligent transportation systems (ITS) have been investigated to improve the safety and efficiency of the transportation industry. Along this line, IEEE developed a protocol called IEEE 802.11p~\cite{Shah2018}. However, this standard is not sufficiently competent to meet the real-time requirements of the emerging vehicular industry. Consequently, existing studies have emphasized the merits of aerial computing for efficient ITS \cite{hu2021uav, cao2022toward} to achieve reliable and low-latency, smart, and seamless vehicle-to-everything communication. Along with benefits, several research challenges still exist in aerial vehicular computing, which can be overcome by leveraging emerging technologies, such as IRS and AI. For example, \cite{hu2021uav} showed that AI-enabled LAPs help improve the service performance of vehicular networks compared with ones without AI and LAP deployment.

\begin{figure}[t]
\centering
\includegraphics[width=1.0\linewidth]{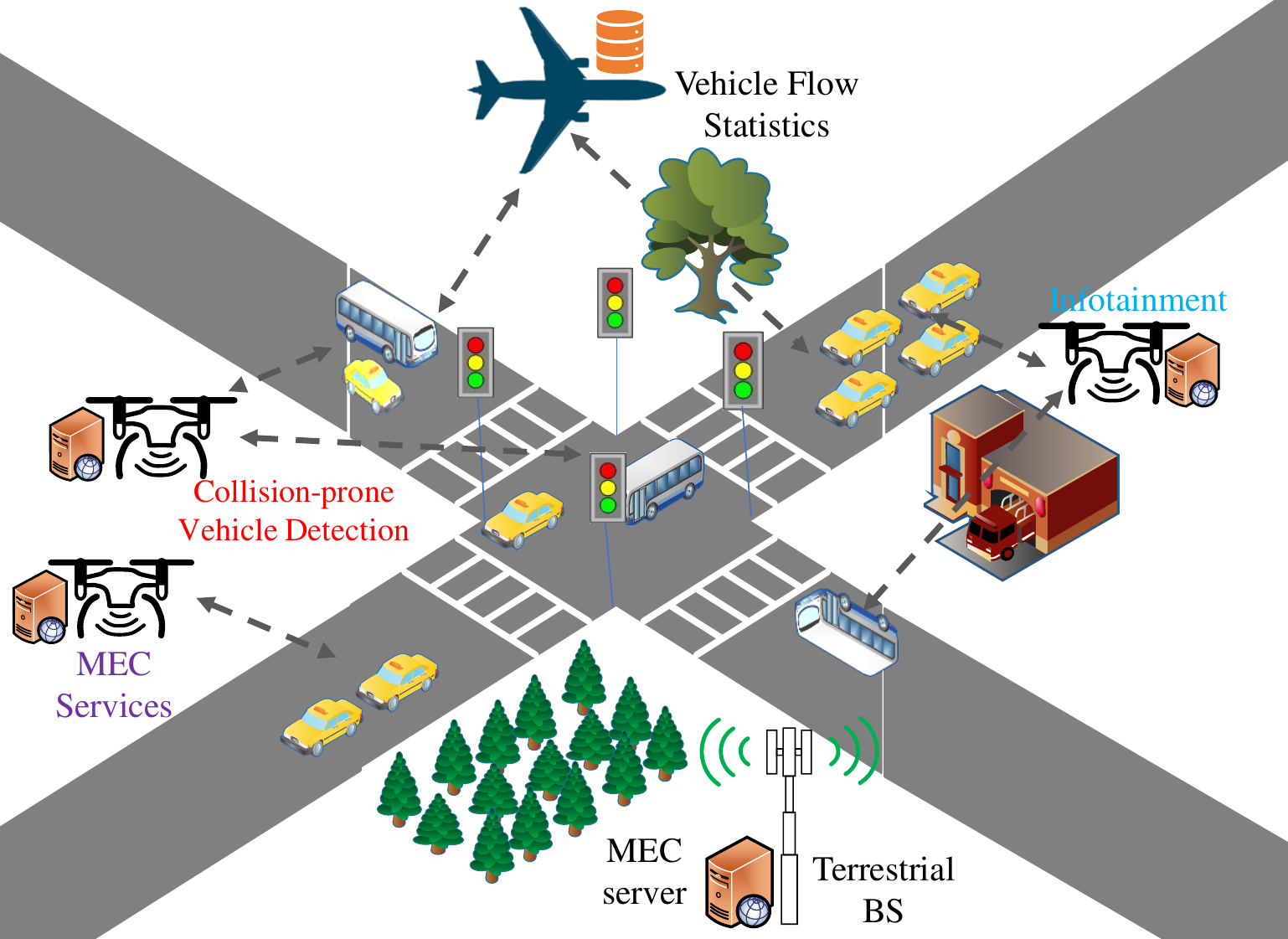}
\caption{Illustration of smart vehicles using aerial computing.} 
\label{Fig:smart-vehicle}
\end{figure}

Furthermore, in recent years, the 5G Automotive Association (5GAA) defined a new concept, namely cooperative intelligent transportation systems (C-ITS), which considers aerial computing as an enabling technology for V2X communications~\cite{sharma2020}. According to this standard, we highlight a few main use cases in the context of C-ITS, such as traffic management, safety and security, MEC provision, and infotainment. 
Fig.~\ref{Fig:smart-vehicle} illustrates a few applications served by aerial computing platforms in smart vehicles. 

\subsubsection{Traffic Management} 
With the growing number of vehicles in recent years, vehicle detection has become a vital part of the ITS, which assists in resolving various information and management issues, such as traffic and vehicle flow statistics and road network management. By acquiring high-resolution images using HACs, the authors of \cite{ZChen2016} designed a vehicle detection algorithm (shown in Fig.~\ref{Fig:smart-vehicle}). Implemented at the HAC server, the detection method considers sparsity in data as well as multiple features. Further, through experiments, the authors verified the utility of the proposed algorithm in terms of detecting a large number of vehicles in metropolitan cities.

\subsubsection{Safety and Security} 
Based on the data collected by the LAC server, the authors of \cite{DRoy2019} proposed the Siamese Interaction Long Short-Term Memory (SILSTM) network to identify the behavior of collision-prone vehicles. Upon learning the interaction scenarios of a vehicle with its neighbors, the SILSTM network can predict accident-prone interactions in advance. Subsequently, through extensive experiments with a real dataset, the authors verified the merit of the proposed approach in terms of detecting collision-prone trajectories at four different intersections. A collision-prone scenario and its resolution are illustrated in the intersection of Fig.~\ref{Fig:smart-vehicle}. Traffic surveillance is one of the core components of next-generation ITS. LACs or HACs can act as relays to facilitate quick data delivery between vehicles and edge nodes, which is associated with a high risk of information leakage. To this end, for the LAC server, the authors of \cite{SGarg2018} proposed an optimization model based on a probabilistic data structure (PDS)-based strategy with a triple Bloom filter that can detect cyber threats of smart vehicles in advance. 

\subsubsection{MEC Support} 
During unexpected events, e.g., adverse weather conditions and extreme traffic congestion, LAC/HACs can act as flying RSUs. Such a scenario was described in \cite{HEl-Sayed2019}, providing the required infrastructure for diverse traffic applications for the improvement of the QoS (shown in Fig.~\ref{Fig:smart-vehicle}). Through simulation, the authors demonstrated that the proposed method could achieve satisfactory performance in terms of network coverage and latency. In the context of social Internet of Vehicles (SIoV) to strengthen social relationship among vehicles, an LAC platform for SIoV, consisting of a three-layer integrated architecture, was adopted in \cite{LZhang2018}. The aim of the architecture was to jointly optimize the transmission power of vehicles and the LAC trajectory under realistic constraints. 
Through simulations compared with existing schemes, the authors demonstrated that the proposed architecture in the LAC platform could effectively maintain social relationships among vehicles.

\subsubsection{Infotainment} 
To disseminate information (e.g., news and entertainment), \cite{SOrtiz2019} proposed a new LAC-based wireless access infrastructure using the RaptorQ-protected content diffusion technique. A simple illustration of this function is shown at the right-hand side of Fig.~\ref{Fig:smart-vehicle}. Through experiments using real vehicles and LAC platforms, the authors demonstrated that the RaptorQ-based content dissemination mechanism in LAC platforms was highly efficient and cost-effective in terms of transmitting information to multiple moving receivers simultaneously. For the same purpose, in \cite{SAHadiwardoyo2019}, a 3D LAC-based simulation model was constructed using the OMNET++ simulator with the objective of passing information from the LACs to cars. Various path loss and elevation models were considered to render the simulation more realistic.  

Disseminating data efficiently is one of the core tasks in V2X networks. To this end, the authors of \cite{RLu2019} proposed a novel LAC-based scheduling protocol that consists of a proactive caching policy and a file-sharing strategy. Subsequently, through simulations, the authors demonstrated that the proposed scheduling strategy could enhance the performance of the system in terms of throughput and latency. Further, to enhance the security of V2X communications, in \cite{BShang2019}, a few physical layer security (PLS) strategies were proposed for LACs and HACs. Comprehensive simulations were conducted, the outcome of which revealed that the proposed strategies in LACs and HACs could effectively handle many emerging security threats in V2X communications.  

\subsection{Smart Factory}
\begin{figure}[t]
\centering
\includegraphics[width=1.0\linewidth]{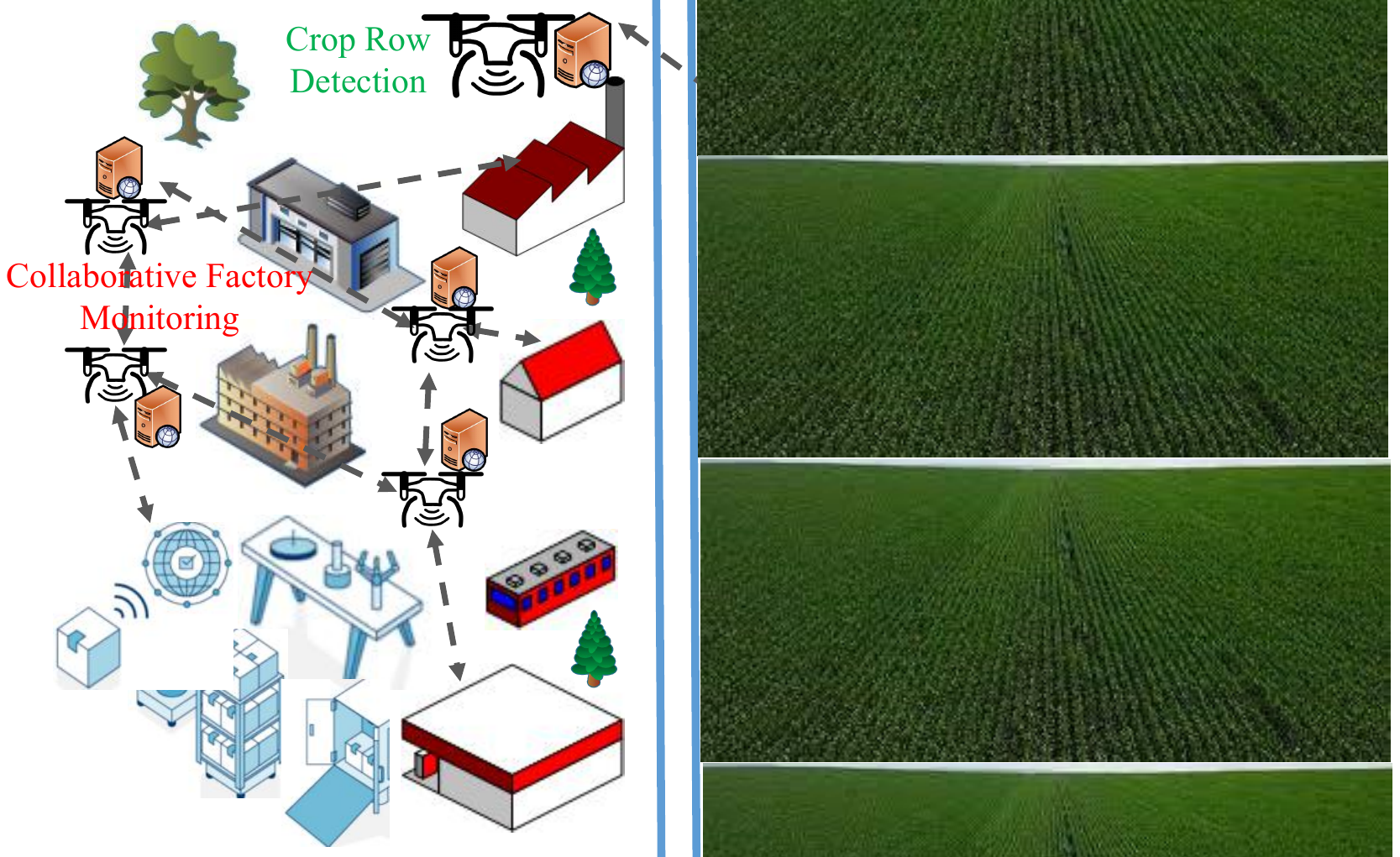}
\caption{Illustration of smart factories using aerial computing.} 
\label{Fig:smart-factory}
\end{figure}

Among the many enabling technologies, 6G, industrial IoT (IIoT) and aerial computing have been integrated with cognitive skills and innovation to aid industries in increasing production and delivering customized products more quickly~\cite{Julian2020}. Such enabling technologies constitute Industry 5.0, an advanced production model focusing on the interaction between machines and humans. This collaborative work facilitates human capabilities, realizing higher productivity and exceptionally easy automation for individuals and small businesses. Such a feat was not possible a few years back. Here, we provide a few examples of manufacturing models with the assistance of LACs and HACs, the corresponding schematic of which is illustrated in Fig.~\ref{Fig:smart-factory}.

With the increasing trend of automation, IIoT has become prevalent in smart factories for data communication and production management. Wireless sensor networks (WSNs), the most important components of the IIoT, are used for numerous tasks in indoor industrial environments, such as irrigation and the inspection of machines. However, these networks have a low computational capability. Therefore, in \cite{HLi2018}, a WSN-based safe navigation algorithm was proposed for micro aerial robots (i.e., LACs) in the IIoT. A WSN with a 3D ranger was used as a computational tool for micro flying robots to detect obstacles in an indoor industrial environment, as shown in Fig.~\ref{Fig:smart-factory}. Through extensive simulations with multiple micro flying robots, the authors verified the merit of the proposed algorithm in terms of obstacle-aware monitoring in dynamic factory environments. Meanwhile, agriculture is becoming increasingly automated with the development of robots and smart tractors. However, to guide these machines in an automated manner, the accurate detection of crop rows is crucial. In \cite{MDBah2020}, based on images acquired by LACs, the authors proposed a learning CRowNet model to detect crop rows (shown in Fig.~\ref{Fig:smart-factory}). The corresponding LAC server was equipped with a model formed using SegNet (S-SegNet) and a CNN-based Hough transform (HoughCNet) technique. Through extensive numerical simulations compared with traditional methods, the authors demonstrated that the CRowNet model achieved 93.58\% accuracy. 

\subsection{Smart Grid}
Smart grids have been widely used in recent years. A large number of new power lines have been under real operation, resulting in a significant growth of transmission line mileage and power equipment. However, the monitoring, detection, and maintenance of power transmission systems are essential to ensure uninterrupted operations. Having noticed the risk of failure in conventional manual monitoring methods, many researchers have been working on the automation of power line inspection, monitoring, and management using the latest technologies~\cite{Yang2020, Meloni2017}, such as aerial computing, image processing, AI, and DL. Most of the existing studies emphasize two crucial aspects of smart grid automation: 1) monitoring and inspection and 2) control and management. We focus on the computational aspect of aerial technologies in performing these tasks. In Fig.~\ref{Fig:smart-gird}, we briefly illustrate a few use cases of the smart grid domain using aerial computing.

\subsubsection{Monitoring and Inspection}
\begin{figure}[t]
\centering
\includegraphics[width=1.0\linewidth]{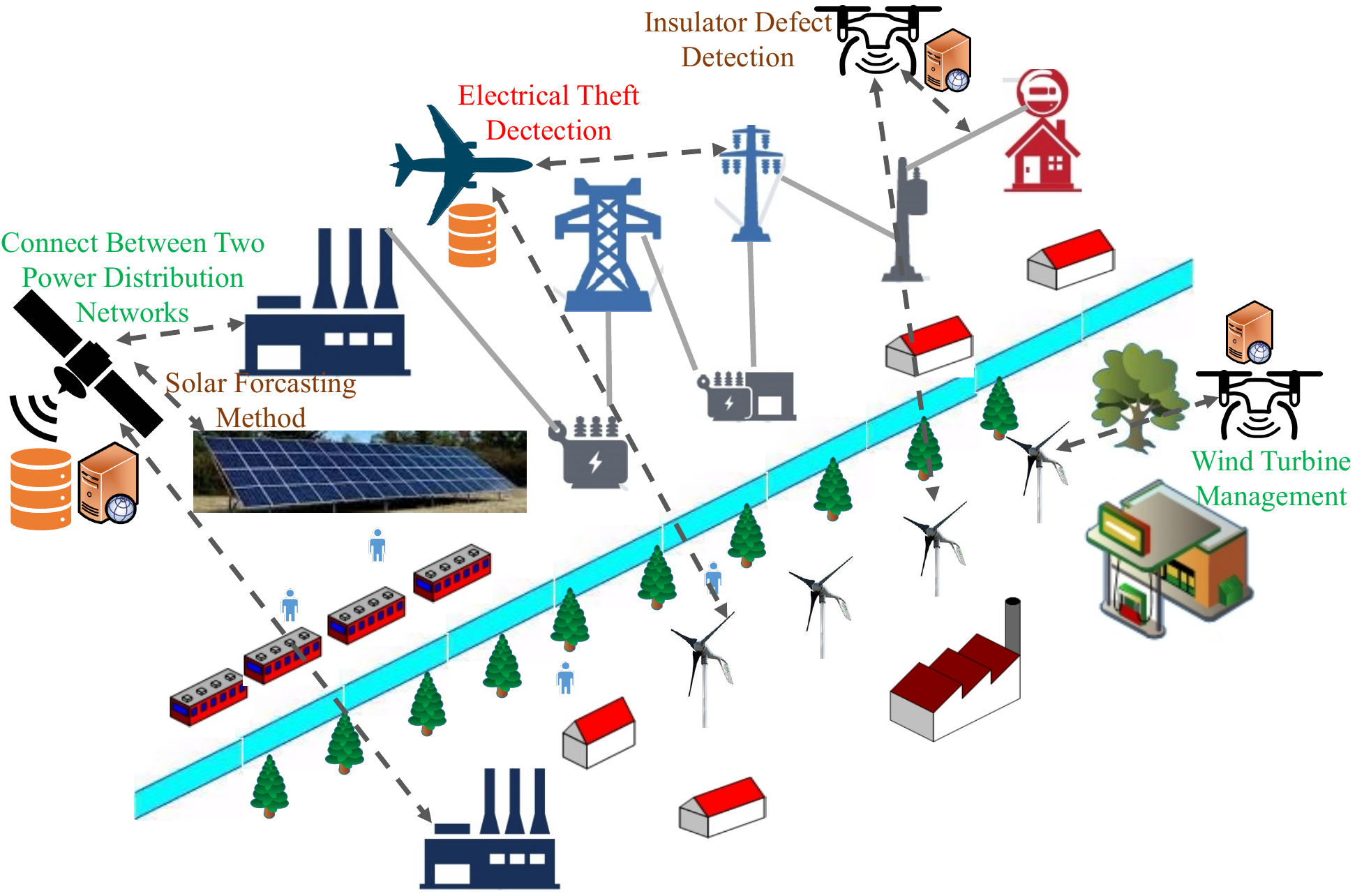}
\caption{Illustration of smart grid using aerial computing.} 
\label{Fig:smart-gird}
\end{figure}

Insulators are prevalent in high-voltage transmission lines and play a key role in electrical insulation and conductor conjunction. Insulator faults (e.g., glass insulators self-blast) pose a grave threat to power systems as they can cause cascade failure. Considering the difficulty of manual inspection, \cite{ZLing2019} proposed a new DL framework to detect the location of broken insulators by capturing images using HACs (shown in Fig.~\ref{Fig:smart-gird}). Specifically, the proposed DL model in the HAC server works in the low-SNR regime with the assistance of two modules: R-CNN to detect objects and U-net to classify pixels. Using a real dataset, via experiments compared with the other methods, the authors showed that the proposed approach achieves real-time accuracy. After capturing images of transmission lines using HACs, \cite{XTao2020} also proposed an improved insulator defect detection method based on an improved ResNeSt and region proposal network. Through suitable experiments on a real dataset, the authors showed that the proposed method deployed in the HAC server can achieve 98.38\% accuracy in detecting insulator defects. Further, aerial computing is a feasible solution to inspect power transmission lines. Consequently, while solving the battery limitation problem of aerial computing platforms (e.g., HACs and LACs), a new idea was proposed in \cite{ZLiu2019}, suggesting the use of a smart hangar as an assistant to achieve full power automation, a path planning model for the LAC server was developed. Through simulations, the authors demonstrated that the proposed scheme in the LAC server is effective in solving existing inspection problems. 

With the growing popularity of wind power, more turbines are being deployed in remote areas. \cite{HMChung2020} considered the deployment of HACs and LACs for inspecting turbines (shown in Fig.~\ref{Fig:smart-gird}). Using the forecasted wind conditions obtained by the HAC-borne edge intelligence, two algorithms were developed to minimize the flight time of the HAC and to optimize power generation by the turbine. Through simulation using a real dataset, it was shown that the proposed method can generate 44\% more power and reduce the flight time by 25\% compared to an existing scheme. With the growth of rooftop solar PV arrays in power generation, \cite{li2020solarfinder} designed a new system called SolarFinder, which can be implemented in a satellite computing platform to detect distributed solar PV arrays over an area in a cost-effective and autonomous manner, as shown in Fig.~\ref{Fig:smart-gird}. The intelligence of SolarFinder is built on the support vector machine (SVM) model and a deep CNN approach jointly. Over 269,632 images captured by satellite, the performance of SolarFinder was evaluated, and the results showed that the pre-trained SolarFinder achieved a Matthews correlation coefficient (MCC) of 0.17.

\subsubsection{Control and Management}
According to the requirements of a smart grid network, controlling non-technical losses (e.g., electrical theft) is one of the most alarming concerns in developing countries. In \cite{PRengaraju2014}, the authors discussed a possible communication network and a framework using LAC platforms to control non-technical losses (i.e., tampering of electrical devices) in energy distribution systems. An example of this use case is shown in Fig.~\ref{Fig:smart-gird}. For the effective management of power distribution networks, which consist of numerous sparsely distributed grids, LEO satellites were utilized as the key components of the communication infrastructure to manage active networks \cite{QYang2012}. Such a scenario is shown at the left corner of Fig.~\ref{Fig:smart-gird}. The effectiveness of the proposed LEO satellite-equipped framework in terms of managing future power supply networks was established through extensive experiments. In \cite{Santics2016}, the feasibility of satellite communication for connecting a smart grid with remotely located actuators was also studied. In addition to the routing and channel access mechanism, the possible optimization of computation resources was studied. In \cite{ZSi2021}, images were captured using a satellite, and a hybrid solar forecasting method was proposed to run the smart grid in a cost-effective manner. For the core computing unit, desirable features and cloud cover factors were extracted from the images using certain intermediate mathematical formulae and a modified CNN, respectively. Coupled with meteorological information and cloud cover factors, a final forecasting model was developed.

\begin{table*}[!ht]
	\centering
	\caption{Summary of vertical domain applications.}
	\label{Table:Vertical-Domain}
	\resizebox{\textwidth}{!}{
	\begin{tabular}{|p{0.23cm}|p{1.6cm}|p{4.4cm}|p{3.75cm}|p{5.1cm}|}
        \cline{2-5}
        \multicolumn{1}{c|}{} & 
        \textbf{References} &
        \textbf{Use Cases} & 
        \textbf{Analytical Tools} & 
        \textbf{Relevant Lessons} \textbf{}
        \\
		\hline
		\multirow{7}{*}{\rotatebox[origin=c]{90}{\textbf{Smart Cities~}}}
		 &  \textbf{--} \cite{YLiu2021, GZhu2019, CHLiu2019, WChen2019, TLi2019, GXu2018,GXu2018-2} \newline
\textbf{--} \cite{JWang2020, XDiao2019, XDiao2019-2, Ztan2020} \newline
\textbf{--} \cite{XLiu2021, MWan2019, PLi2019, ZHu2019,JGao2021, Colli2020, LMa2019, LLiu2020,YZang2016, KKuru2021, YChen2019, Gong2021}\newline
\textbf{--} \cite{EErtugrul2018, SHu2020, YJin2020, HKim2018, GKVerma2020, Keke2020,aloqaily2021design}
		 &	    \textbf{--} Data collection, high-mobility multimodal sensing, control and management of sensing, enhancement of QoS in crowd sensing, improvement of information coverage, and tracking moving targets.\newline
                 \textbf{--} Serving IoT devices with efficient resource utilization, task data allocation, and trajectory optimization with/without NOMA, and serving as cache-enabled edge computing nodes. \newline 
                  \textbf{--} Automatic extraction of building instances, target tracking, monitoring urban environments, air quality monitoring, rainfall monitoring, semi-automated driving line automation, building road networks, city planning, extraction of urban tree inventory, and modeling an urban parking lot. \newline 
                 \textbf{--} Physical security of smart buildings, monitoring suspicious links, mitigating challenges of video surveillance applications, constructing surveillance models for multi-domain IoT, ensuring security in data dissemination, and trustworthy authentication.
		&
              
              \textbf{--} K-means method, Hungarian algorithm, differential geometry, CNN and DDPG, edge/cloud computing, IRHS, sparse algorithms based on compressive sensing technology, and a novel parametric sparse imaging approach. \newline
              \textbf{--} Geographic merging of tasks, optimization algorithms, and convex approximation. \newline
              -- Sparse representation theory, collaborative workload sharing algorithm, system development, placement optimization, satellite microwave downlink sensor networks, inverse distance weighted method, a novel ADSM technique, decentralized agent-based control architecture, classification-based method, and SLAM algorithm. \newline
              \textbf{--} Jamming signal, an optimal scheduling algorithm, optimization, low-cost CB-PS scheme, and blockchain.

        &

            \textbf{--} Minimize the total transmission power of IoT nodes, outperforms three existing schemes, improves QoS, improves the coverage ratio of information by 21.42\%, and can track moving targets accurately. \newline
            \textbf{--} Provide energy-efficient AR services to mobile users under tight delay constraints. \newline
            \textbf{--} Enhance performance compared with the mainstream Mask R-CNN method; achieve good precision compared with state-of-the-art works; maximize the total weight of completed tasks; achieve a lower AQI recognition error; has close agreement with other existing methods; achieve an average recall, precision, and F1-score of 90.79\%, 92.94\%, and 91.85\%, respectively; achieve an overall accuracy of 78\%; and achieve centimeter-level accuracy with a precision of 84.8\%. \newline
            \textbf{--} Achieve an acceptable accuracy for indoor environments, effectively monitor a suspicious link, make video surveillance applications work efficiently, enhance the security of the LAC server, perform authentication tasks effectively, and proves that blockchain is effective in collaborative security among multiple LACs and HACs. \\
                
		\hline 
		\multirow{7}{*}{\rotatebox[origin=c]{90}{\textbf{Smart Vehicles~}}}
		 & \textbf{--} \cite{ZChen2016} \newline
\textbf{--} \cite{DRoy2019, SGarg2018} \newline
\textbf{--} \cite{HEl-Sayed2019, LZhang2018} \newline
\textbf{--} \cite{SOrtiz2019,SAHadiwardoyo2019, RLu2019}
		 &	
              \textbf{--} Vehicle detection for traffic management and road network planning. \newline
              \textbf{--} Traffic collision detection and cyber threat detection. \newline
              \textbf{--} Flying roadside units, and social relationships among vehicles. \newline
              \textbf{--} Disseminating news and entertainment information among vehicles and sharing files in V2X. 
              
		&
		      \textbf{--} ML technique. \newline
              \textbf{--} SILSTM network and a PDS-based scheduling technique. \newline
              \textbf{--} Algorithm and optimization tools. \newline
              \textbf{--} RaptorQ-based content dissemination mechanisms and optimized scheduling policies.

        &
            \textbf{--} Achieve satisfactory performance in vehicle detection. \newline
            \textbf{--} Can detect collision-prone interaction trajectories and cyber threats effectively. \newline
            \textbf{--} Can achieve full network coverage under different scenarios and optimize the transmit power in vehicles and LAC trajectory. \newline
            \textbf{--} Can deliver infotainment service effectively. \\
		\hline 
	
	    \multirow{4}{*}{\rotatebox[origin=c]{90}{\textbf{Factories}}}
	    & \textbf{--} \cite{HLi2018} \newline
\textbf{--} \cite{MDBah2020}

		 &	
		     \textbf{--} Monitoring industrial environments. \newline
             \textbf{--} Detecting crop rows in agriculture.
        &
              \textbf{--} A navigation algorithm. \newline
              \textbf{--} A model formed with SegNet (S-SegNet) and HoughCNet.

        &
            \textbf{--} Can navigate accurately while detecting obstacles. \newline
            \textbf{--} Can detect crop row with the accuracy of 93.58\%. 
                \\
		\hline 
	
	    \multirow{7}{*}{\rotatebox[origin=c]{90}{\textbf{Smart Grids~}}}
	    & \textbf{--} \cite{ZLing2019,XTao2020, ZLiu2019, HMChung2020, li2020solarfinder} \newline
\textbf{--} \cite{PRengaraju2014, QYang2012, Santics2016, ZSi2021}

		 &	
		     \textbf{--} Detecting insulator defects in transmission lines, inspecting power transmission lines, inspecting wind turbines, and detecting distributed solar PV arrays. \newline
             \textbf{--} Controlling non-technical losses in energy distribution systems, managing active power distribution networks, connecting two power distribution networks, and forecasting solar energy for the smooth operation of smart grids.
             
		&
            \textbf{--} DL and customized neural networks, optimization tools, and integration of an SVM model with deep CNNs. \newline
	        \textbf{--} Design of a suitable communication network; integration of LEO satellites, optimization tools, and CNN-based approaches. 

      &
            \textbf{--} Achieve 93.58\% accuracy, achieve 44\% improvement in power generation and 25\% reduction in flight time, achieve an MCC of $0.17$, which is $3$ times better than a pre-trained CNN-based approach, proves that a smart hanger can effectively solve the limited battery capacity problem of LACs/HACs. \newline
            \textbf{--} Can detect tampering of electrical devices, show that LEO satellites are effective in managing remote power generation sites, and can forecast solar energy for smooth operation. 
                \\
		\hline
	\end{tabular}
	}
\end{table*}

\subsection{Summary and Discussion}
In this section, important vertical domain applications supported by aerial computing, including smart cities, smart vehicles, smart factories, and smart grids, have been discussed. A summary of the applications is presented in Table~\ref{Table:Vertical-Domain}. 

Data communication, processing, and management of ground sensors and IoT nodes are crucial tasks to facilitate applications in smart cities. 
Aerial computing platforms (e.g., LACs, HACs, and satellites) are effective in performing these tasks owing to their flexible deployment and LoS communication features in air-to-air or air-to-ground links. However, even with the use of aerial computing, several research challenges require to be solved to obtain optimal performance, such as 3D deployment, joint computation and communication resource management, coverage, and trajectory design. Several existing studies have addressed these challenges by adopting cutting-edge techniques, such as convex optimization, graph theory, and AI/ML tools. Because of security breaches, smart city applications are associated with several challenges, such as designing low-cost and efficient signature and authentication protocols. Few studies have adopted the CB-PS scheme and blockchain to solve these problems to a certain extent. Further, aerial computing is a promising solution for realizing MEC with other communication technologies, which is associated with several challenges as well, such as task data allocation, computation and communication resource management, and aerial device deployment issues. City planning and management are also crucial aspects of smart cities. Certain useful tasks in the context of this use case are target tracking to facilitate city design, air pollution monitoring, and city map designs. In addition to providing certain functional requirements, aerial computing may aid in achieving surveillance and security in smart cities. The effective utilization of aerial computing is associated with certain challenges, such as the HAC/LAC navigation strategy, deployment, and resource management. 

In the domain of smart vehicles, aerial computing platforms can be effectively utilized for various purposes, such as traffic management, safety and security, flying RSUs, and infotainment. Despite serving all these use cases, several research challenges need to be solved, such as computation and communication resource management, trajectory design of aerial computing platforms, data analysis, and routing strategy. Several existing studies have addressed these issues via AI/ML techniques, probabilistic approaches, and optimization tools.

Finally, monitoring, inspection, and management are the primary applications of aerial computing platforms in smart manufacturing and smart grid domains. However, certain technical problems must be solved to obtain optimal outcome from aerial computing platforms in these domains, such as obstacle-aware monitoring a factory environment \cite{HLi2018} and crop row detection (using ML tools) in agriculture field \cite{MDBah2020}. Furthermore, a few more problems in smart grids using aerial computing, such as the detection of insulator defects, management of solar PV array, and inspection of power line and wind turbine, need to be resolved. 
In terms of control and management, the main problems are electrical theft detection, active and power distribution network management, and forecasting solar energy for the smooth and continued operation of smart grids, which can be addressed by equipping special hardware and software in aerial computing platforms, AI/ML, and optimization tools.

\section{Challenges and Future Directions}
\label{sec:Challenges}
In this section, we identify interesting research challenges and highlight possible future directions for aerial computing. 

\subsection{Energy Efficiency }
In aerial computing, achieving sustainable energy management at LAC and HAC platforms and energy-efficient satellite computing is a major concern. Data communications and service delivery in satellite environments and space travel with airplanes, require large energy resources to ensure network operations. 
In addition, the power consumption of aerial components may vary depending on the operation mode. For example, average power consumption of the UAV is $8.2412$ W in the idle mode and $245.2815$ W in the horizontal movement mode, and average power consumption is $8.2618$ W and $8.2637$ W when the UAV performs communication transmissions with WiFi and GPS, respectively \cite{abeywickrama2018empirical}. Thus, we need to focus on energy-efficient designs to realize sustainable and green aerial computing. For example, the authors of \cite{Challenge6} suggested a space-air-ground architecture with a focus on maximizing the system energy efficiency enabled by the joint optimization of uplink transmission power control, sub-channel selection, and deployment of aerial relays. This can be achieved by dividing the original problem into two sub-problems, optimal subchannel selection and power control policy, which are obtained by available aerial relay deployment. Moreover, energy refilling techniques to exploit renewable energy resources can be useful for building sustainable aerial computing systems. Aerial devices and satellites can harvest power from ambient environments, such as wind, solar, vibration, and thermal power, at the LAC platform to support their operations, e.g., data communications over aerial links.

\subsection{Resource Management} 
Aerial computing accommodates data collection and computation tasks to serve end IoT devices via different computing platforms (e.g., LAC and HAC) and supports edge services in future 6G networks. Compared to terrestrial computing infrastructures, such as cloud data centers and cloudlet, computing platforms in aerial computing possess limited storage resources and battery capabilities, which would hinder the deployment of aerial computing services, such as airplane-based civil monitoring for terrestrial IoT environments \cite{Challenge1}. Thus, resource management strategies are urgently required to satisfy seamless service delivery in space. Moreover, in the future 6G era, intelligent aerial computing is expected to be a dominant research area, where AI functions can be integrated into aerial devices at the LAC platform to enable self-controlled and autonomous aerial systems. In this context, training DL models on these aerial devices with massive datasets may be infeasible because of the high demands of computation and memory resources, particularly when training with large-scale audio and image data \cite{Challenge2}. Therefore, optimizing on-device AI/DL models is of paramount importance for solving the computational burden posed on aerial devices in the air. Consequently, several solutions have been proposed to facilitate resource management in aerial computing. For example, the authors of \cite{Challenge3} considered a resource management solution for aerial computing systems wherein a joint optimization of task offloading, resource allocation mechanism, and the trajectory of aerial devices in the air was derived with respect to terrestrial users' latency requirements. Thus, the energy of the aerial devices was minimized, while a longer flight duration was achieved. Moreover, it is also necessary to develop on-device learning solutions to support self-learning aerial computing systems. For example, improved network architecture, training optimization, and hardware design were used to accelerate on-device data training \cite{zhou2021device}. 
A streamlined slimming framework was developed and combined with a consecutive tensor layer to improve the training rates. Simulation results show that the proposed method can enhance the training rate by up to 30\% compared to traditional approaches without compromising learning accuracy. Thus, this approach provides opportunities for designing intelligent flying devices in aerial computing.

\subsection{Network Stability} 
The topology of aerial computing fluctuates with the number of nodes, battery levels, and varying communication conditions. The trajectories and speeds of aerial components and flying devices also vary owing to the terrestrial application requirements and environmental dynamics, e.g., different altitudes of buildings in smart cities in the LAC platform and different orbital altitudes of multiple LEO satellites. Further, the operational complexity and unpredictability of aerial components in space can make the involved aerial computing system unstable. In fact, aerial nodes can join and leave unpredictably, or their flight can be stationary, slow, or fast. Thus, achieving network sustainability in aerial systems is critical. As a promising method, an optimal tracking policy was constructed in \cite{Challenge9} for each aerial device in an aerial-based network to mitigate the varying network topology issues in aerial computing operations, e.g., flying trajectories of UAVs. The key focus was on achieving an adaptive surrounding network configuration for varying channel quality and communication bandwidth resources. Subsequently, a particle swarm optimization algorithm was developed to optimally schedule the energy allocation among a set of aerial nodes, while the prediction error of the surrounding node locations was minimized. Another study in \cite{Challenge10} built an observer, which could monitor surrounding unmanned aerial nodes in an aerial environment using Kalman filtering with respect to the maximum number of parallel targets, measurement time, measurement success rate, and measurement noise. Thus, the optimal measurement policy was obtained for network topology monitoring, aiming to achieve an accurate prediction of trajectory topology in autonomous aerial computing systems. Moreover, blockchain can help build scalable aerial computing networks thanks to its decentralized feature for enabling large-scale interconnection among servers and devices, e.g., unmanned aerial components in the HAC platform \cite{ghribi2020secure}. Each aerial device works as a blockchain node to build decentralized data services, such as data sharing, data communications, and data storage, without the need for a central server for better networking scalability.

\subsection{Large-Scale Network Optimization} 
In aerial computing, aerial devices operate on a large scale, and cooperative optimization is needed to utilize the advantages from multiple and distributed network datasets, such as diverse channel features and environment properties, while improving the network quality. For example, performing an optimal trajectory control policy for all manned and unmanned aerial components is a challenge on the HAC platform if only the characteristics of an aerial device are obtained \cite{Challenge11}. More importantly, the datasets of future intelligent aerial networks are distributed over large-scale networks rather than being centrally located. Therefore, there is an urgent need for distributed and large-scale optimization approaches to enable scalable and intelligent aerial computing applications. A large-scale trajectory optimization solution was proposed in \cite{Challenge12} for the Internet of aerial devices, for example, UAVs and balloons, through the use of a multi-agent DRL algorithm. This enables aerial devices to collaboratively develop a distributed sense-and-send protocol in an aerial computing setting, facilitating large-scale sensing and data task transfer in cellular networks. 

\subsection{Security, Privacy, and Trust} 
Although 6G-based aerial computing can offer global coverage and diverse QoS provision to industrial applications, critical issues related to security, privacy, and trust must be solved. Adversaries may attack aerial communication channels and deploy data breaches in the flying BSs on the HAC platform, as the management of lower altitude-based servers is limited owing to the physical distance \cite{Challenge14}. AI techniques are extensively used to enable intelligent aerial computing, but they often require centralized data collection for training, which raises potential privacy issues owing to the exposure of sensitive information in the air. Moreover, the deployment of satellite-terrestrial communications in space can face critical challenges caused by untrusted environments, as third parties and attackers can compromise the data exchange among aerial devices, BSs, and terrestrial IoT users.  Therefore, blockchain is a promising solution \cite{Challenge15} for building trust and establishing secure decentralized satellite-ground communications for aerial computing systems. Blockchain, with its traceability and trustworthiness, can provide enhanced security for aerial computing systems \cite{lei2019securing}. The data from aerial communications, such as device location and channel information, can be stored in the data ledger where blocks are built based on immutable transactions. This makes the information in the blockchains resistant to data modifications or threats. The elimination of centralized authority of the blockchain also helps mitigate single-point failures in involved aerial networks. In addition, smart contracts can be used as self-executing software running on blockchain for providing automatic authentication and verification to ensure reliable aerial communications \cite{Challenge15}. This technique attains further relevance in the 6G era because aerial computing systems tend to be decentralized and deployed on a large scale, which can be realized using the decentralization feature of blockchain. To preserve data privacy in aerial computing, perturbation techniques, such as differential privacy and dummy operations, are particularly helpful in protecting data leakage against external threats during data exchange. For example, differential privacy was adopted in \cite{wang2021learning} to realize privacy-enhanced intelligent aerial communications by introducing artificial noise into the trained gradients at aerial devices. This solution helps hide sensitive information while guaranteeing convergence, such that adversaries cannot retrieve useful data samples. 

\section{Conclusion}
\label{sec:Conclusion}
Edge computing has become an indispensable component of the present network infrastructure, but it also has various limitations owing to the emergence of new services and applications and the expansion of the network. We highlighted the fact that the current computing infrastructure does not meet new demands and requirements. In this regard, the concept of \textit{aerial computing} was introduced and comprehensively reviewed in this study. 
First, we presented an overview of aerial computing, starting from the system architecture to the reference model and fundamental features, and compared it with conventional computing paradigms, such as cloudlet, fog computing, and MEC. Next, we presented a set of key enabling technologies and discussed the use of aerial computing in vertical domain applications. Finally, we discussed the research challenges and promising future directions pertaining to aerial computing. As the development of aerial computing is still in a preliminary stage and there are many unexplored issues, we believe that this paper has revealed certain important lessons and key ideas that will drive further research and unlock the full potential of a comprehensive 6G computing infrastructure in the future.

\balance

\end{document}